\documentclass[twocolumn,balancelastpage,pra,showpacs]{revtex4}
\usepackage{amsmath}
\usepackage{graphicx}

\input{tcilatex}

\begin{document}

\title{Theory of quantum fluctuations of optical dissipative structures and
its application to the squeezing properties of bright cavity solitons}
\author{Isabel P\'{e}rez-Arjona$^{\left( 1\right) }$, Eugenio Rold\'{a}n$%
^{\left( 2\right) }$, and Germ\'{a}n J. de Valc\'{a}rcel$^{\left( 2\right) }$%
}
\affiliation{$^{\left( 1\right) }$Departament de F\'{\i}sica Aplicada, Escola Polit\`{e}%
cnica Superior de Gandia, Universitat Polit\`{e}cnica de Val\`{e}ncia, Ctra.
Natzaret-Oliva s/n, Spain}
\affiliation{$^{\left( 2\right) }$Departament d'\`{O}ptica, Universitat de Val\`{e}ncia,
Dr. Moliner 50, 46100-Burjassot, Spain}

\begin{abstract}
We present a method for the study of quantum fluctuations of dissipative
structures forming in nonlinear optical cavities, which we illustrate in the
case of a degenerate, type I optical parametric oscillator. The method
consists in (i) taking into account explicitly, through a collective
variable description, the drift of the dissipative structure caused by the
quantum noise, and (ii) expanding the remaining -internal- fluctuations in
the biorthonormal basis associated to the linear operator governing the
evolution of fluctuations in the linearized Langevin equations. We obtain
general expressions for the squeezing and intensity fluctuations spectra.
Then we theoretically study the squeezing properties of a special
dissipative structure, namely, the bright cavity soliton. After reviewing
our previous result that in the linear approximation there is a perfectly
squeezed mode irrespectively of the values of the system parameters, we
consider squeezing at the bifurcation points, and the squeezing detection
with a plane--wave local oscillator field, taking also into account the
effect of the detector size on the level of detectable squeezing.
\end{abstract}

\maketitle

\section{Introduction}

Quantum fluctuations in the light field observables have been a subject of
intensive and maintained research since the appearance of quantum optics in
the early sixties, especially since the discovery of the nonclassical states
of light and the generation of squeezed light in the mid eighties \cite%
{Drummond04}. Quantum fluctuations are unavoidable as their origin lies in
the impossibility of determining two canonically conjugate observables with
a precision better than that allowed by the Heisenberg uncertainty
relations. In the case of the radiation field, quantum fluctuations manifest
in quantities such as the photon number, the field phase, the Stokes
parameters, or the field quadratures \cite%
{Loudon87,Meystre91,Korolkova02,Dodonov03,Walls94}.

Although its origin could be traced back till the early days of quantum
mechanics \cite{Dodonov03}, the modern squeezing research goes back to the
mid seventies, its foundational period being closed with the experimental
generation of squeezed light in the mid eighties \cite{Meystre91}. In a
single--mode state the two field quadratures (equivalent to the position and
momentum operators of a harmonic oscillator) constitute a canonically
conjugate pair, the product of their uncertainties being consequently
limited by the Heisenberg inequality. When the field mode is in a coherent
state, the uncertainty is the same for any field quadrature and equals that
of the vacuum state. Then, a squeezed state is that for which the
uncertainty in a particular field quadrature is smaller than that of the
quantum vacuum, a reduction achieved at the expense of an increase in the
uncertainty of the complementary field quadrature. These states can be
generated in nonlinear processes (four-wave mixing, parametric
down-conversion...), and the field quadratures can be detected in a homodyne
detection experiment in which the quantum state is mixed with a classical
(intense, coherent) local oscillator field \cite{Loudon87}.

Single-mode squeezing has been intensively and extensively investigated for
almost three decades \cite{Drummond04}. A well established result is that a
high degree of squeezing is obtained in nonlinear cavities near the
bifurcation points, and that the squeezing level degrades as the system is
brought far from them \cite{Walls94}. Multimode squeezing has also been
considered in the past. One of the most immediate cases is that of nonlinear
cavities working in several longitudinal modes, a type of system already
exhibiting noise reduction features specific to multimode systems. For
example, the quantum noise suppression on the difference of intensities
(amplitude-squeezing) of a two--mode optical parametric oscillator above
threshold \cite{Reynaud87,Reid88}, which is associated to the existence of a
continuous diffusion of the phase difference between the two modes. It is to
be remarked that squeezing appears in this case associated to a \textit{%
collective variable}, namely the intensity difference, and that, remarkably,
the noise reduction level is independent of the system proximity to a
bifurcation point.

A particularly interesting case of multimode squeezing is that of solitons.
Squeezing in optical fibre solitons (\textit{temporal} solitons) has
attracted much interest since the late eighties \cite{Drummond87,Haus90} and
is by now quite well understood, see e.g. \cite{Kozlov03} for a recent
review. In this problem the relevance of collective variables, such as the
position or the momentum of the soliton, is very clear in the sense that
these are the observables in which the behavior of quantum fluctuations is
easily detectable. As we comment below, this problem has some similarities,
but also strong differences,\ with the problem of cavity solitons that we
treat here. A related problem, that of the squeezing of spatial solitons in
Kerr media, has also been considered recently \cite%
{Treps00,Lantz04,Nagasako98,Mecozzi98}.

For our purposes, a most exciting connection is that existing between
pattern formation in nonlinear optical cavities and squeezing \cite%
{Lugiato99,Lugiato02}. In these systems the concepts used in the analysis of
quantum fluctuations, which has started relatively recently, must be
generalized to cover correlations at different spatial points \cite%
{Kolobov99}.

Extended nonlinear optical systems, specially nonlinear optical cavities
with large Fresnel numbers, are systems that spontaneously display
dissipative structures, which are extended patterns that form in the plane
orthogonal to the light propagation direction, through an spontaneous
symmetry breaking. One of the new concepts that appeared when the analysis
of quantum fluctuations in these systems was addressed is the quantum image 
\cite{Lugiato95}, which can be described as a precursor appearing below
threshold of the pattern that the system would display above threshold. The
quantum image is not detectable at low observation frequencies as in this
case the fast dynamics of quantum fluctuations is washed out. We refer the
reader to existing reviews for a resume of the main researches in the field 
\cite{Lugiato99,Lugiato02}. For our purposes, which concern the analysis of
the squeezing in dissipative structures, the work carried out by Gatti and
Lugiato \cite{Lugiato93,Gatti95} on the squeezing of extended degenerate
optical parametric oscillators (DOPO) below threshold, is particularly close
(see also \cite{Drummond05}), as well as the analysis of the role played by
the Goldstone mode in \cite{Zambrini00,Gomila02}.

In this article we investigate quantum fluctuations of dissipative
structures in the DOPO. In a first part we develop the theory for the
calculation of linearized quantum fluctuations, in particular for the
calculation of the squeezing and intensity fluctuations spectra. The theory
is general in the sense that no particular dissipative structure is assumed.
Moreover, although the model for a DOPO in the large pump detuning is used,
the derived expressions are easily generalizable to cover other nonlinear
cavity models. Then, in a second part we apply this theory to the study of
the squeezing properties of a special dissipative structure appearing above
threshold, the so--called bright cavity soliton.

In \cite{EPL} we already advanced a particular property of quantum
fluctuations which is specific to pattern formation (i.e., that is absent
when the emission is homogeneous in space), which can be put in short as
follows:\ Due to the translational invariance of the problem (the position
of the pattern in the transverse plane is not fixed when the pumping is
spatially homogeneous), there is a particular transverse mode that is free
from quantum fluctuations (in the linear approximation), a \textit{perfectly}
squeezed\textit{\ }mode. And this occurs irrespectively of the value of the
parameters of the system. The only conditions are that (i) the system be
translational invariant, and (ii)\ that the output field displays a pattern.
This particular mode corresponds to the transverse linear momentum of the
pattern.

Here we shall go beyond this result by studying the squeezing properties of
a special pattern, the bright cavity soliton. For that we shall consider the
DOPO\ model in the large pump detuning limit, as the problem is much simpler
in these conditions because there exists an explicit analytical solution for
the localized structure. As stated, the article consists of two parts. In
the first part (covering Sections II to V) we derive general expressions for
the study of quantum fluctuations (e.g., the linear squeezing spectrum)
without specializing to any particular pattern. Then, the second part
(Section VI) is devoted to the bright cavity soliton. Finally, in Section
VII we give our main conclusions.

\section{Linear theory of quantum fluctuations of optical dissipative
structures}

In this section we present the general method that will allow us determining
quantities related with the quantum fluctuations of optical dissipative
structures (e.g. the squeezing spectrum) in the linear approximation. An
outstanding feature of the method is that it circumvents the numerical
integration of the dynamical, stochastic equations, which is always a
problematic (and terribly time consuming) task. Instead, the method exploits
the diagonalization of the linear problem, which allows simplest formal
solutions.

We shall use the DOPO\ with plane cavity mirrors as a model for presenting
the method. Any other nonlinear optical cavity with plane mirrors sustaining
dissipative structures can be studied along similar ways after
straightforward particularization of the expressions given below.

We assume that the dynamical equations of the studied system have been cast
in the form of classical-looking Langevin equations corresponding to some
coherent state representation. In particular we assume that a generalized $P$
representation \cite{Drummond80} is being used (Sec. II.A) as its normal
ordering equivalence allows a direct computation of measurable quantities
corresponding to the fields leaving the cavity. Furthermore we assume that
those Langevin equations have been linearized around a classical dissipative
structure (Sec. II.B). The method consists in separating the fluctuations in
two classes: (i) Those coming from the drift of the dissipative structure,
as it can move freely across the transverse section owed to the spatial
translation invariance of the model (Sec. II.C); and (ii) Formally solving
the equations corresponding to the remaining ("internal") fluctuations
making use of a special basis (Sec. II.D), what allows deriving a general
expression for the linearized squeezing spectrum (Sec. III), as well as for
the spectrum of intensity fluctuations (Sec. IV). This section ends with a
general result concerning the squeezing of dissipative structures (Sec. V).

\subsection{\label{DOPOmodel}Langevin equations for a planar DOPO in the
generalized P representation}

We consider the model for a type I DOPO\ with plane cavity mirrors of \cite%
{Gatti97} pumped by a plane wave coherent field of frequency $2\omega_{%
\mathrm{s}}$ and amplitude $\mathcal{E}_{\mathrm{in}}$. An intracavity $%
\chi^{(2)}$ nonlinear crystal converts pump photons into subharmonic
photons, at frequency $\omega_{\mathrm{s}}$, and vice versa. Only two
longitudinal cavity modes, of frequencies $\omega_{0}$ (pump mode) and $%
\omega_{1}$(signal mode), which are the closest to $2\omega_{\mathrm{s}}$
and $\omega_{\mathrm{s}}$, respectively, are assumed to be relevant. The
cavity is assumed to be single-ended, i.e., losses occur at a single cavity
mirror, where the intracavity modes are damped at rates $\gamma_{n}$, $n=0,1$%
. The above frequencies define dimensionless pump and signal detuning
parameters through $\Delta_{0}=\left( \omega_{0}-2\omega_{\mathrm{s}}\right)
/\gamma_{0}$ and $\Delta_{1}=\left( \omega_{1}-\omega_{\mathrm{s}}\right)
/\gamma_{1}$, respectively.

We denote the intracavity field envelope operators for pump and signal
modes, which propagate along the $z$ direction, by $\hat{A}_{0}\left( 
\mathbf{r},t\right) $ and $\hat{A}_{1}\left( \mathbf{r},t\right) $,
respectively, where\textbf{\ }$\mathbf{r}=\left( x,y\right) $ is the
transverse position vector, obeying standard equal-time commutation relations%
\begin{equation}
\left[ \hat{A}_{m}\left( \mathbf{r},t\right) ,\hat{A}_{n}^{\dag}\left( 
\mathbf{r}^{\prime},t\right) \right] =\delta_{m,n}\delta\left( \mathbf{r}-%
\mathbf{r}^{\prime}\right) .  \label{commutationrel}
\end{equation}

As commented we shall use a coherent state representation in order to handle
the problem. These representations set a correspondence between the quantum
operators $\hat{A}_{m}(\mathbf{r},t)$ and $\hat{A}_{m}^{\dag }(\mathbf{r},t)$
and the c-number fields $\mathcal{A}_{m}(\mathbf{r},t)$ and $\mathcal{A}%
_{m}^{+}(\mathbf{r},t)$, respectively, which are independent but in their
(stochastic) averages, which verify $\left\langle \mathcal{A}_{m}^{+}(%
\mathbf{r},t)\right\rangle =\left\langle \mathcal{A}_{m}(\mathbf{r}%
,t)\right\rangle ^{\ast }$. The physical meaning of the stochastic average
of any function of the c-number fields depends on the representation. In
particular we shall use the generalized $P$ representation \cite{Drummond80}%
, generalized to include the spatial nature of the multimode problem here
considered \cite{Gatti97}, in which the stochastic averages correspond to
quantum expectation values computed in normal and time ordering.

As shown in Appendix A, in the large pump detuning limit ($\left\vert \Delta
_{0}\right\vert \gg 1,\left\vert \Delta _{1}\right\vert ,\gamma _{0}/\gamma
_{1}$) the DOPO dynamical (Langevin) equations in the generalized $P$
representation can be written as

\begin{subequations}
\label{AdiabaticLangevin}
\begin{gather}
\frac{\partial }{\partial t}\mathcal{A}_{1}(\mathbf{r},t)=\gamma _{1}\left(
L_{1}\mathcal{A}_{1}+\mu \mathcal{A}_{1}^{+}+i\frac{\sigma }{\kappa ^{2}}%
\mathcal{A}_{1}^{2}\mathcal{A}_{1}^{+}\right) +  \notag \\
\sqrt{\gamma _{1}\left( \mu +i\frac{\sigma }{\kappa ^{2}}\mathcal{A}%
_{1}^{2}\right) }\eta (\mathbf{r},t), \\
\frac{\partial }{\partial t}\mathcal{A}_{1}^{+}(\mathbf{r},t)=\gamma
_{1}\left( L_{1}^{\ast }\mathcal{A}_{1}^{+}+\mu \mathcal{A}_{1}-i\frac{%
\sigma }{\kappa ^{2}}\mathcal{A}_{1}^{+}{}^{2}\mathcal{A}_{1}\right) + 
\notag \\
\sqrt{\gamma _{1}\left( \mu -i\frac{\sigma }{\kappa ^{2}}\mathcal{A}%
_{1}^{+}{}^{2}\right) }\eta ^{+}(\mathbf{r},t),
\end{gather}%
where $L_{1}=-(1+i\Delta _{1})+il_{1}^{2}\nabla ^{2}$, $l_{1}=c/\sqrt{%
2\omega _{1}\gamma _{1}}$ is the diffraction length for the signal field, $%
\nabla ^{2}=\partial ^{2}/\partial x^{2}+\partial ^{2}/\partial y^{2}$ is
the transverse Laplacian operator, and we have introduced the real and
dimensionless parametric pump parameter $\mu $%
\end{subequations}
\begin{equation}
\mu =\frac{g\left\vert \mathcal{E}_{\mathrm{in}}\right\vert }{\gamma
_{0}\gamma _{1}\left\vert \Delta _{0}\right\vert }>0,
\end{equation}%
where $g$ is the (real) nonlinear coupling coefficient, Eq. (\ref{defg}) in
Appendix A, the normalized nonlinear coupling coefficient%
\begin{equation}
\kappa ^{-2}=\frac{g^{2}}{2\gamma _{0}\gamma _{1}\left\vert \Delta
_{0}\right\vert },
\end{equation}%
and $\sigma =\func{sign}\Delta _{0}=\pm 1$. Finally, $\eta (\mathbf{r},t)$
and $\eta ^{+}(\mathbf{r},t)$ are independent, real white Gaussian noises of
zero average and correlations given by Eqs. (\ref{noise}) in Appendix A.

Equations (\ref{AdiabaticLangevin}) are the model we shall consider along
this paper.

\subsection{\label{linlan}Dynamics of quantum fluctuations: Linearized
Langevin equations around the classical dissipative structures of the DOPO}

In the classical limit ($\mathcal{A}_{i}^{+}$ being interpreted as $\mathcal{%
A}_{i}^{\ast }$, noises being ignored), Eqs. (\ref{AdiabaticLangevinClassic}%
) have the form of a parametrically driven nonlinear Schr\"{o}dinger
equation (PDNLSE), first derived for the DOPO in \cite{Longhi97,Trillo97},
see Eq. (\ref{PDNLSE}) in Appendix B. In our case $\sigma $ accounts for the
cases of self-focusing $\left( \sigma =+1\right) $ or defocusing $\left(
\sigma =-1\right) $ of the PDNLSE, which determine the kind of dissipative
structures (DS) supported by the DOPO. These DS are patterns that appear in
the transverse plane with respect to the direction of light propagation. We
denote these (steady) structures by $\mathcal{A}_{1}\left( \mathbf{r}\right)
=\mathcal{\bar{A}}_{1}\left( \mathbf{r}-\mathbf{r}_{1}\right) $, where $%
\mathbf{r}_{1}=\left( x_{1},y_{1}\right) $ is arbitrary due to the
translation invariance of the problem. They can be, e. g., periodic patterns
or localized structures \cite%
{Longhi95,Longhi97,Bondila95,Barashenkov02,deValcarcel02}. Although in the
second part of this paper we shall concentrate on a particular type of DS,
namely the bright cavity soliton, we stress here that the treatment we
present below is completely general and covers the description of quantum
fluctuations of any stationary DS.

The dynamics of the quantum fluctuations around any DS is studied by setting 
\begin{subequations}
\label{ClassicalSolution}
\begin{align}
\mathcal{A}_{1}(\mathbf{r},t) & =\mathcal{\bar{A}}_{1}\left( \mathbf{r}-%
\mathbf{r}_{1}\left( t\right) \right) +a_{1}\left( \mathbf{r}-\mathbf{r}%
_{1}\left( t\right) ,t\right) , \\
\mathcal{A}_{1}^{+}(\mathbf{r},t) & =\mathcal{\bar{A}}_{1}^{\ast}\left( 
\mathbf{r}-\mathbf{r}_{1}\left( t\right) \right) +a_{1}^{+}\left( \mathbf{r}-%
\mathbf{r}_{1}\left( t\right) ,t\right) ,
\end{align}
where $\mathcal{\bar{A}}_{1}\ $and $\mathcal{\bar{A}}_{1}^{\ast}$ are the
classical stationary mean values of the field corresponding to a particular
DS (i.e., the stationary solutions of Eqs. (\ref{AdiabaticLangevin}) when
the noise terms are neglected), and $a_{1}$ and $a_{1}^{+}$ are the c-number
fields accounting for the quantum fluctuations. Notice that the position of
the classical solution, $\mathbf{r}_{1}\left( t\right) =\left(
x_{1},y_{1}\right) $, is let to vary in time in order to properly describe
the diffusive movement of the DS, which is excited by (quantum) noise.

Linearizing\ the Langevin equations around the classical solution we get, to
first order in the fluctuations, the linearized equation of motion for the
quantum fluctuations, that read 
\end{subequations}
\begin{equation}
-\kappa \left( \mathbf{G}_{x}\frac{\mathrm{d}x_{1}}{\mathrm{d}t}+\mathbf{G}%
_{y}\frac{\mathrm{d}y_{1}}{\mathrm{d}t}\right) +\frac{\partial }{\partial t}%
\mathbf{a}_{1}=\gamma _{1}\mathcal{L}\mathbf{a}_{1}+\sqrt{\gamma _{1}}%
\mathbf{h},  \label{linearizedLangevin}
\end{equation}%
where 
\begin{equation}
\mathbf{G}_{x(y)}=\partial _{x(y)}\left( 
\begin{array}{c}
\mathcal{\bar{A}}_{1} \\ 
\mathcal{\bar{A}}_{1}^{\ast }%
\end{array}%
\right) ,  \label{goldstone}
\end{equation}%
$\mathbf{a}_{1}$ is the quantum fluctuations vector 
\begin{equation}
\mathbf{a}_{1}(\mathbf{r},t)=\left( 
\begin{array}{c}
a_{1}(\mathbf{r},t), \\ 
a_{1}^{+}(\mathbf{r},t)%
\end{array}%
\right) ,
\end{equation}%
$\mathbf{h}$ is the noise vector 
\begin{equation}
\mathbf{h}(\mathbf{r},t)=\left( 
\begin{array}{c}
\sqrt{\bar{\alpha}_{0}}\ \eta (\mathbf{r},t) \\ 
\sqrt{\bar{\alpha}_{0}^{\ast }}\ \eta ^{+}(\mathbf{r},t)%
\end{array}%
\right) ,
\end{equation}%
where%
\begin{equation}
\bar{\alpha}_{0}=\mu +i\sigma \kappa ^{-2}\mathcal{\bar{A}}_{1}^{2}.
\end{equation}%
Finally, the linear operator $\mathcal{L}$ and its adjoint $\mathcal{L}%
^{\dag }$ read 
\begin{subequations}
\label{LinearOperator}
\begin{align}
\mathcal{L}& =\left( 
\begin{array}{cc}
\mathcal{L}_{1} & \bar{\alpha}_{0} \\ 
\bar{\alpha}_{0}^{\ast } & \mathcal{L}_{1}^{\ast }%
\end{array}%
\right) ,\ \ \mathcal{L}^{\dag }=\left( 
\begin{array}{cc}
\mathcal{L}_{1}^{\ast } & \bar{\alpha}_{0} \\ 
\bar{\alpha}_{0}^{\ast } & \mathcal{L}_{1}%
\end{array}%
\right) , \\
\mathcal{L}_{1}& =-(1+i\Delta _{1})+il_{1}^{2}\nabla ^{2}+2i\sigma \kappa
^{-2}\left\vert \mathcal{\bar{A}}_{1}(\mathbf{r})\right\vert ^{2},
\label{Laux}
\end{align}%
where we note that a typo has been corrected in Eq. (\ref{Laux}) with
respect to the corresponding expression in \cite{EPL}.

In Eq. (\ref{linearizedLangevin}) two terms ($\partial _{x}\mathbf{a}_{1}%
\mathrm{d}x_{1}/\mathrm{d}t$ and $\partial _{y}\mathbf{a}_{1}\mathrm{d}y_{1}/%
\mathrm{d}t$) have been neglected as they are of second order in the
fluctuations. Notice finally that Eq. (\ref{linearizedLangevin}) has the
standard form of a set of linearized Langevin equations but for the first
term appearing on the left hand side, which describes possible displacements
of the DS on the transverse plane.

All the information about quantum fluctuations in the linear approximation
is contained in Eq. (\ref{linearizedLangevin}). Our goal is then finding an
efficient method for solving it at the time that relevant information is
extracted in a transparent way. This is accomplished by using the
eigensystems of the linear operators

\subsection{Diagonalization of the linear problem. The role of the Goldstone
modes: Drift of the dissipative structure}

Our main purpose in this work is to study the properties of quantum
fluctuations around the semiclassical mean value corresponding to a DS. In
this section we solve Eqs. (\ref{linearizedLangevin}), which\ will allow to
characterize the quantum fluctuations, in particular through the squeezing
spectrum. With this aim, we develop a general method suitable for obtaining
the formal solution of Eqs. (\ref{linearizedLangevin}), suitable for any
system and any classical stationary DS.

Let us assume without proof that the set of eigenvectors of the linear
operators $\mathcal{L}$ and $\mathcal{L}^{\dag}$, Eq. (\ref{LinearOperator}%
), form a biorthonormal basis. (In the second part of this article we show
that this is indeed the case for the bright cavity soliton solution, unlike
the problem of conservative temporal solitons where the set of eigenvectors
must be completed in order to form a proper basis \cite{Kozlov03}). The
method used to solve Eq. (\ref{linearizedLangevin}) consists in expanding
the quantum fluctuations in this biorthonormal basis.

We denote the eigensystems of $\mathcal{L}$ and $\mathcal{L}^{\dag }$ by 
\end{subequations}
\begin{subequations}
\label{Eigensystem}
\begin{align}
\mathcal{L}\mathbf{v}_{i}(\mathbf{r})& =\lambda _{i}\mathbf{v}_{i}(\mathbf{r}%
),\ \ \ \mathbf{v}_{i}(\mathbf{r})=%
\begin{pmatrix}
v_{i}(\mathbf{r}) \\ 
v_{i}^{+}(\mathbf{r})%
\end{pmatrix}%
,  \label{Ldiag} \\
\mathcal{L}^{\dag }\mathbf{w}_{i}(\mathbf{r})& =\lambda _{i}^{\ast }\mathbf{w%
}_{i}(\mathbf{r}),\ \ \mathbf{w}_{i}(\mathbf{r})=%
\begin{pmatrix}
w_{i}(\mathbf{r}) \\ 
w_{i}^{+}(\mathbf{r})%
\end{pmatrix}%
.  \label{Ldaggerdiag}
\end{align}

In the above and in the following expressions, the index $i$ represents both
the discrete and the continuous spectra as we do not want to overburden the
notation. Note also that, in the following, Kronecker deltas should be
understood as suitable Dirac deltas as well as sums should be understood as
suitable integrals when referring to the continuous part of the spectra.

We define scalar product as usual 
\end{subequations}
\begin{equation}
\left\langle \mathbf{u}|\mathbf{s}\right\rangle =\int\mathrm{d}^{2}r~\mathbf{%
u}^{\dag}(\mathbf{r})\cdot\mathbf{s}(\mathbf{r}),  \label{ScalarProduct}
\end{equation}
so that relation 
\begin{equation}
\left\langle \mathbf{w}_{i}|\mathcal{L}\mathbf{s}\right\rangle =\lambda
_{i}\left\langle \mathbf{w}_{i}|\mathbf{s}\right\rangle ,
\end{equation}
holds for any $\mathbf{s}$. Finally, all eigenvectors are assumed to be
normalized as%
\begin{equation}
\left\langle \mathbf{w}_{i}|\mathbf{v}_{j}\right\rangle =\delta_{ij}.
\end{equation}

The spectra must be computed numerically in general. However, it will be
convenient for our purposes to state two general properties of the discrete
spectra. These are:

\begin{itemize}
\item Property 1: For any parameter set there exist Goldstone modes, $%
\mathbf{v}_{1x(1y)}=\mathbf{G}_{x(y)}$, with $\mathbf{G}_{x(y)}$ given by
Eq. (\ref{goldstone}), satisfying 
\begin{equation}
\mathcal{L}\mathbf{v}_{1x(1y)}=0,
\end{equation}
and the associated adjoint eigenvectors, denoted by $\mathbf{w}_{1x(1y)}$,
which verify 
\begin{equation}
\mathcal{L}^{\dag}\mathbf{w}_{1x(1y)}=0.
\end{equation}
This property is a consequence of the translational invariance of the
problem, as any DS can be located at any position on the transverse plane.

\item Property 2: For any parameter set there exist eigenvectors of the
adjoint problem, which we denote as $\mathbf{w}_{2x(2y)}$, verifying 
\begin{equation}
\mathcal{L}^{\dag }\mathbf{w}_{2x(2y)}=-2\mathbf{w}_{2x(2y)}.
\end{equation}%
These eigenvectors are%
\begin{equation}
\mathbf{w}_{2x(2y)}(\mathbf{r})=\partial _{x(y)}\left( 
\begin{array}{c}
i\mathcal{\bar{A}}_{1}(\mathbf{r}) \\ 
-i\mathcal{\bar{A}}_{1}^{\ast }(\mathbf{r})%
\end{array}%
\right) ,  \label{lofmagic}
\end{equation}%
as is straightforward to be checked. This property will be associated to a
perfectly squeezed mode \cite{EPL}, as we show below.
\end{itemize}

Now, the linearized Langevin equation (\ref{linearizedLangevin}) can be
solved by expanding the quantum fluctuations on the basis $\left\{ \mathbf{v}%
_{i}\right\} $,%
\begin{equation}
\mathbf{a}_{1}(\mathbf{r},t)=\sum\limits_{i}c_{i}(t)\mathbf{v}_{i}(\mathbf{r}%
),  \label{expansion}
\end{equation}%
where the Goldstone modes have been excluded from this expansion as any
contribution of them to $\mathbf{a}_{1}(\mathbf{r},t)$ would entail a shift
of the solution, which is already accounted for by the (still undetermined)
location of the DS, by definition.

First we project Eq. (\ref{linearizedLangevin}) onto $\mathbf{w}_{1x}$ and $%
\mathbf{w}_{1y}$, obtaining 
\begin{subequations}
\label{dif}
\begin{align}
\dot{x}_{1} & =-\frac{\sqrt{\gamma_{1}}}{\kappa}\left\langle \mathbf{w}_{1x}|%
\mathbf{h}\right\rangle ,  \label{dif1} \\
\dot{y}_{1} & =-\frac{\sqrt{\gamma_{1}}}{\kappa}\left\langle \mathbf{w}_{1y}|%
\mathbf{h}\right\rangle ,  \label{dif2}
\end{align}
We see that the DS diffuses driven by quantum fluctuations as anticipated.
The diffusive drift of the DS through a time $t_{\mathrm{d}}$ can be
evaluated by considering the mean squared deviation of the position of the
DS along this time, 
\end{subequations}
\begin{equation}
\boldsymbol{\rho}(t)=\mathbf{r}_{1}(t)-\mathbf{r}_{1}(t-t_{\mathrm{d}}).
\label{drift}
\end{equation}
The variance $\left\langle \boldsymbol{\rho}^{2}(t)\right\rangle $ can be
calculated from Eqs.(\ref{dif}), which give the time evolution for $\mathbf{r%
}_{1}=(x_{1},y_{1})$, obtaining%
\begin{equation}
x_{1}(t)=-\frac{\sqrt{\gamma_{1}}}{\kappa}\int\limits_{0}^{t}\mathrm{d}%
t^{\prime}\left\langle \mathbf{w}_{1x}\left( \mathbf{r}\right) |\mathbf{h}%
\left( \mathbf{r},t^{\prime}\right) \right\rangle ,  \label{x1}
\end{equation}
and analogous expression holds for $y_{1}(t)$, when $\mathbf{w}_{1x}$ is
replaced by $\mathbf{w}_{1y}$ in (\ref{x1}). Substituting this solution into
Eq.(\ref{drift}), we reach the expression of the variance of $\boldsymbol{%
\rho }(t)$, which is linear in time \cite{Gomila02}, and reads%
\begin{equation}
\left\langle \boldsymbol{\rho}^{2}(t)\right\rangle =Dt_{\mathrm{d}},
\label{variance}
\end{equation}
where the diffusion coefficient is given by 
\begin{equation}
D=2\frac{\gamma_{1}}{\kappa^{2}}\func{Re}\int\mathrm{d}^{2}r\left[
w_{1x}^{2}(\mathbf{r})+w_{1y}^{2}(\mathbf{r})\right] \bar{\alpha}_{0}^{\ast
}(\mathbf{r}).  \label{diffconstant}
\end{equation}
The knowledge of the variance (\ref{variance}) allows us to evaluate the
possible effects of the DS movement on the noise detected, e.g., in a
homodyning experiment. This could be quantitatively important as a possible
noise reduction could be blurred by the diffusion of the structure \cite{EPL}%
.

\subsection{Formal solution to the linearized Langevin equations}

Now by substituting (\ref{expansion}) into (\ref{linearizedLangevin}) and
projecting onto $\left\{ \mathbf{w}_{i}\right\} $, we obtain the evolution
equation for the expansion coefficients in Eq. (\ref{expansion}):%
\begin{equation}
\dot{c}_{i}=\gamma_{1}\lambda_{i}c_{i}+\sqrt{\gamma_{1}}\left\langle \mathbf{%
w}_{i}|\mathbf{h}\right\rangle ,  \label{CoefficientEvolution}
\end{equation}
where the index $i$ does not include the Goldstone modes, as commented. We
write down Eq. (\ref{CoefficientEvolution}) in the temporal-frequencies
space. By defining the Fourier transforms 
\begin{subequations}
\label{FourierTransf}
\begin{align}
c_{i}(t) & =\frac{1}{2\pi}\int\mathrm{d}\omega~e^{i\omega t}\tilde{c}%
_{i}(\omega), \\
\mathbf{h}(\mathbf{r},t) & =\frac{1}{2\pi}\int\mathrm{d}\omega~e^{i\omega t}%
\mathbf{\tilde{h}}(\mathbf{r},\omega),
\end{align}
and the corresponding inverse transforms 
\end{subequations}
\begin{subequations}
\label{InverseFourier}
\begin{align}
\tilde{c}_{i}(\omega) & =\int\mathrm{d}t~e^{-i\omega t}c_{i}(t), \\
\mathbf{\tilde{h}}(\mathbf{r},\omega) & =\int\mathrm{d}t~e^{-i\omega t}%
\mathbf{h}(\mathbf{r},t),
\end{align}
we find a simple expression in the\ temporal spectral domain for Eq.(\ref%
{CoefficientEvolution}), that reads 
\end{subequations}
\begin{equation}
i\omega\tilde{c}_{i}(\omega)=\gamma_{1}\lambda_{i}\tilde{c}_{i}(\omega )+%
\sqrt{\gamma_{1}}\left\langle \mathbf{w}_{i}\left( \mathbf{r}\right) |%
\mathbf{\tilde{h}}\left( \mathbf{r},\omega\right) \right\rangle ,
\end{equation}
which gives 
\begin{equation}
\tilde{c}_{i}(\omega)=\frac{\sqrt{\gamma_{1}}\left\langle \mathbf{w}%
_{i}\left( \mathbf{r}\right) |\mathbf{\tilde{h}}\left( \mathbf{r}%
,\omega\right) \right\rangle }{i\omega-\gamma_{1}\lambda_{i}}.
\label{CoefOmega}
\end{equation}

By using solution (\ref{CoefOmega}) in Eqs. (\ref{FourierTransf}), we
retrieve the expansion coefficients $c_{i}(t)$%
\begin{equation}
c_{i}(t)=\frac{\sqrt{\gamma_{1}}}{2\pi}\int\mathrm{d}\omega~e^{i\omega t}%
\frac{\left\langle \mathbf{w}_{i}\left( \mathbf{r}\right) |\mathbf{\tilde {h}%
}\left( \mathbf{r},\omega\right) \right\rangle }{i\omega-\gamma
_{1}\lambda_{i}},  \label{CoefficientExpression}
\end{equation}
which is the formal solution of the linearized Langevin equations, Eqs. (\ref%
{expansion}) and (\ref{CoefficientEvolution}) .

\subsection{Modal correlation spectrum}

Once the time-dependent expansion coefficients $c_{i}(t)$ are known, Eq. (%
\ref{CoefficientExpression}), we can calculate the two--time correlations
between the different modes. The knowledge of these correlations, and
specifically, of their spectra, is necessary in order to characterize some
properties of quantum fluctuations, such as the spectrum of squeezing or the
spectrum of intensity fluctuations of the quantum field exiting the
nonlinear cavity, as will be shown below.

We define the correlation spectrum between two modes labeled by indices $i$
and $j$ as usual%
\begin{equation}
S_{ij}(\omega )=\int \mathrm{d}\tau ~e^{-i\omega \tau }\left\langle
c_{i}(t+\tau ),c_{j}(\tau )\right\rangle  \label{ModalSpectrumDef}
\end{equation}%
where the correlation%
\begin{multline}
\left\langle c_{i}(t+\tau ),c_{j}(\tau )\right\rangle = \\
\left\langle c_{i}(t+\tau )c_{j}(\tau )\right\rangle -\left\langle
c_{i}(t)\right\rangle \left\langle c_{j}(t)\right\rangle ,
\end{multline}%
and $\left\langle \mathcal{O}\right\rangle $ is the (stochastic) average
value of $\mathcal{O}$. By using Eqs. (\ref{InverseFourier}) and\ (\ref%
{CoefOmega}), $S_{ij}(\omega )$ can be written as 
\begin{multline}
S_{ij}(\omega )=\frac{\gamma _{1}}{2\pi }\int \mathrm{d}\omega ^{\prime }%
\frac{e^{i(\omega ^{\prime }+\omega )\tau }}{\left( \gamma _{1}\lambda
_{i}-i\omega \right) \left( \gamma _{1}\lambda _{j}-i\omega ^{\prime
}\right) }\times \\
\left\langle \left\langle \mathbf{w}_{i}|\mathbf{\tilde{h}}(\mathbf{r}%
,\omega )\right\rangle ,\left\langle \mathbf{w}_{j}|\mathbf{\tilde{h}}(%
\mathbf{r},\omega )\right\rangle \right\rangle .
\end{multline}%
Taking into account the properties of noise, Eqs. (\ref{noise}), one obtains
straightforwardly 
\begin{multline}
\left\langle \left\langle \mathbf{w}_{i}|\mathbf{\tilde{h}}(\mathbf{r}%
,\omega )\right\rangle ,\left\langle \mathbf{w}_{j}|\mathbf{\tilde{h}}(%
\mathbf{r}^{\prime },\omega ^{\prime })\right\rangle \right\rangle = \\
2\pi \delta \left( \omega +\omega ^{\prime }\right) \int \mathrm{d}%
^{2}r~d_{ij}\left( \mathbf{r},\mathbf{r}\right) ,
\end{multline}%
where%
\begin{multline}
d_{ij}\left( \mathbf{r},\mathbf{r}\right) =  \label{dij} \\
w_{i}^{\ast }\left( \mathbf{r}\right) w_{j}^{\ast }\left( \mathbf{r}\right) 
\bar{\alpha}_{0}\left( \mathbf{r}\right) +\left[ w_{i}^{+}\left( \mathbf{r}%
\right) \right] ^{\ast }\left[ w_{j}^{+}\left( \mathbf{r}\right) \right]
^{\ast }\bar{\alpha}_{0}^{\ast }\left( \mathbf{r}\right) ,
\end{multline}%
so that the modal correlation spectrum $S_{ij}(\omega )$ can be written as%
\begin{equation}
S_{ij}(\omega )=\frac{D_{ij}}{\left( \lambda _{i}-i\omega /\gamma
_{1}\right) \left( \lambda _{j}+i\omega /\gamma _{1}\right) },
\label{ModalSpectrum}
\end{equation}%
where we have introduced the matrix $D_{ij}$, which we call modal diffusion
matrix (because of the similarity of Eq. (\ref{ModalSpectrum}) with an
spectral matrix \cite{Collet85}), defined by%
\begin{equation}
D_{ij}=\int \mathrm{d}^{2}r\ d_{ij}\left( \mathbf{r},\mathbf{r}\right) .
\label{diffusionmatrix}
\end{equation}

Note that all modal correlations can be obtained just by diagonalizing the
linear problem.

\section{Squeezing spectrum in the linear approximation}

We consider now the squeezing properties of the classical DS $\mathcal{%
\bar
{A}}_{1}$ as measured via a balanced homodyne detection experiment 
\cite{Gatti95}, which allows the direct measurement of quadrature squeezing.
The quantum field outgoing the nonlinear cavity, to be denoted by $\hat
{A}%
_{1,\mathrm{out}}\left( \mathbf{r},t\right) $, is combined at a beam
splitter with a local oscillator field (LOF). This LOF lies in a classical
multimode coherent state $\mathbf{\alpha}_{\mathrm{L}}\left( \mathbf{r}-%
\mathbf{r}_{\mathrm{L}}\left( t\right) \right) $ of intensity much larger
than that of\ $\hat{A}_{1,\mathrm{out}}\left( \mathbf{r},t\right) $. (The
shift $\mathbf{r}_{\mathrm{L}}\left( t\right) $ is introduced in order to
cover the case of a movable LOF for reasons that will be clear below, such
as to consider the possibility to follow the DS movement.)

The difference $\widehat{\Delta I}(t)$ between the intensities $\mathcal{%
\hat
{D}}_{+}$ and $\mathcal{\hat{D}}_{-}$ of the two output ports of the
beam splitter, with 
\begin{equation}
\mathcal{\hat{D}}_{\pm}(\mathbf{r},t)=\frac{1}{\sqrt{2}}[\hat{A}_{1,\mathrm{%
out}}(\mathbf{r},t)\pm\alpha_{\mathrm{L}}\left( \mathbf{r}-\mathbf{\mathbf{r}%
}_{\mathrm{L}}\left( t\right) \right) ],
\end{equation}
is then measured, and it turns out to be given by \cite{Gatti95} 
\begin{align}
\widehat{\Delta I}(t) & =\int\mathrm{d}^{2}r\left[ \mathcal{\hat{D}}%
_{+}^{\dag}(\mathbf{r},t)\mathcal{\hat{D}}_{+}(\mathbf{r},t)-\mathcal{\hat{D}%
}_{-}^{\dag}(\mathbf{r},t)\mathcal{\hat{D}}_{-}(\mathbf{r},t)\right]  \notag
\\
& \equiv\sqrt{N}\hat{E}_{\mathrm{H},\mathrm{out}}(t).
\end{align}
where the projection of the output signal field $\hat{A}_{1,\mathrm{out}}(%
\mathbf{r},t)$ on the LOF has been introduced as the field $\hat
{E}_{%
\mathrm{H},\mathrm{out}}(t)$ 
\begin{align}
\hat{E}_{\mathrm{H},\mathrm{out}}(t) & =\hat{A}_{\mathrm{H},\mathrm{out}}(t)+%
\hat{A}_{\mathrm{H},\mathrm{out}}^{\dag}(t), \\
\hat{A}_{\mathrm{H},\mathrm{out}}(t) & \equiv\frac{1}{\sqrt{N}}\int\mathrm{d}%
^{2}r~\alpha_{\mathrm{L}}^{\ast}(\mathbf{r}-\mathbf{\mathbf{r}}_{\mathrm{L}%
}\left( t\right) )\hat{A}_{1,\mathrm{out}}(\mathbf{r},t),
\end{align}
with the LOF intensity denoted by%
\begin{equation}
N=\int\mathrm{d}^{2}r\left\vert \alpha_{\mathrm{L}}(\mathbf{r})\right\vert
^{2}.  \label{LOFintens}
\end{equation}

By using (\ref{commutationrel}), the spectrum of the difference intensity
fluctuations can be written as%
\begin{multline}
V(\omega )=\int\limits_{-\infty }^{\infty }\mathrm{d}\tau ~e^{-i\omega \tau
}\left\langle \hat{E}_{\mathrm{H},\mathrm{out}}(t+\tau ),\hat{E}_{\mathrm{H},%
\mathrm{out}}(t)\right\rangle \\
=1+\int\limits_{-\infty }^{\infty }\mathrm{d}\tau ~e^{-i\omega \tau
}\left\langle :\hat{E}_{\mathrm{H},\mathrm{out}}(t+\tau ),\hat{E}_{\mathrm{H}%
,\mathrm{out}}(t):\right\rangle \\
=1+S_{\mathrm{out}}(\omega ),
\end{multline}%
which defines the squeezing spectrum $S_{\mathrm{out}}(\omega )$ of the
field exiting the cavity. When $\hat{A}_{\mathrm{out}}(\mathbf{r},t)$ is in
a coherent state, $V(\omega )=1$ and $S_{\mathrm{out}}(\omega )=0$, which
defines the standard quantum limit. Light is said to be squeezed at a
frequency $\omega _{\mathrm{c}}$ when $S_{\mathrm{out}}(\omega _{\mathrm{c}%
})<0$, and the case of complete noise reduction, or perfect squeezing, at $%
\omega _{\mathrm{c}}$ is signaled by $S_{\mathrm{out}}(\omega _{\mathrm{c}%
})=-1$ as in this case $V(\omega _{\mathrm{c}})=0$.

Now the correlations of the output fields can be written in terms of those
of the intracavity fields by using the input-output formalism \cite{Collet84}%
\begin{align}
\left\langle :\hat{A}_{1,\mathrm{out}}^{\dag}(t),\hat{A}_{1,\mathrm{out}%
}(t^{\prime}):\right\rangle & =2\gamma_{1}\left\langle :\hat{A}_{1}^{\dag
}(t),\hat{A}_{1}(t^{\prime}):\right\rangle  \label{in-out} \\
& =2\gamma_{1}\left\langle \mathcal{A}_{1}^{+}(t),\mathcal{A}_{1}(t^{\prime
})\right\rangle  \notag
\end{align}
and then 
\begin{equation}
S_{\mathrm{out}}(\omega)=2\gamma_{1}\int\limits_{-\infty}^{\infty}\mathrm{d}%
\tau~e^{-i\omega\tau}\left\langle \delta\mathcal{E}_{\mathrm{H}%
}(t+\tau),\delta\mathcal{E}_{\mathrm{H}}(t)\right\rangle ,
\end{equation}
\newline
where $\delta\mathcal{E}_{\mathrm{H}}(t)=\mathcal{E}_{\mathrm{H}%
}(t)-\left\langle \mathcal{E}_{\mathrm{H}}(t)\right\rangle $ with 
\begin{subequations}
\begin{align}
\mathcal{E}_{\mathrm{H}}(t) & =\mathcal{A}_{\mathrm{H}}(t)+\mathcal{A}_{%
\mathrm{H}}^{+}(t), \\
\mathcal{A}_{\mathrm{H}}(t) & =\frac{1}{\sqrt{N}}\int\mathrm{d}^{2}r~\alpha_{%
\mathrm{L}}^{\ast}(\mathbf{r}-\mathbf{r}_{\mathrm{L}}\left( t\right) )%
\mathcal{A}_{1}(\mathbf{r},t).
\end{align}
This expression can be written in a more compact form as 
\end{subequations}
\begin{subequations}
\label{Sout}
\begin{align}
S_{\mathrm{out}}\left( \omega\right) & =\frac{2\gamma_{1}}{N}%
\int\limits_{-\infty}^{\infty}\mathrm{d}\tau e^{-i\omega\tau}\langle \delta%
\mathcal{E}_{\mathrm{H}}\left( t+\tau\right) \delta\mathcal{E}_{\mathrm{H}%
}\left( t\right) \rangle, \\
\delta\mathcal{E}_{\mathrm{H}}\left( t\right) & =\langle\boldsymbol{\alpha }%
_{\mathrm{L}}\left( \mathbf{r}+\boldsymbol{\rho}\left( t\right) \right) \mid%
\mathbf{a}_{1}\left( \mathbf{r},t\right) \rangle,  \label{deltaEh}
\end{align}
where\textbf{\ }$\boldsymbol{\rho}\left( t\right) =$ $\mathbf{r}_{1}\left(
t\right) -\mathbf{r}_{\mathrm{L}}\left( t\right) $, and the change of
variables $\mathbf{r}\rightarrow\mathbf{r}-\mathbf{r}_{1}\left( t\right) $
has been introduced. (Remind that $\mathbf{r}_{1}\left( t\right) $ describes
the position of the dissipative structure, that changes because of the
diffusion introduced by quantum fluctuations, Eqs. (\ref{dif}).)

The output squeezing spectrum (\ref{Sout}) can now be calculated in terms of
the modal correlation spectrum (\ref{ModalSpectrum}): The intracavity field
fluctuations $\mathbf{a}_{1}(\mathbf{r},t)$ can be written in terms of the
expansion (\ref{expansion}) --remind that Goldstone modes are excluded--, so
that the output squeezing spectrum (\ref{Sout}) takes the form 
\end{subequations}
\begin{equation}
S_{\mathrm{out}}(\omega )=\frac{2\gamma _{1}}{N}\sum\limits_{i,j}\left%
\langle \boldsymbol{\alpha }_{\mathrm{L}}|\mathbf{v}_{i}\right\rangle
\left\langle \boldsymbol{\alpha }_{\mathrm{L}}|\mathbf{v}_{j}\right\rangle
S_{ij}(\omega ),  \label{Specsum}
\end{equation}%
where 
\begin{multline}
\left\langle \boldsymbol{\alpha }_{\mathrm{L}}|\mathbf{v}_{i}\right\rangle
=\int\limits_{-\infty }^{\infty }\mathrm{d}^{2}r~\alpha _{\mathrm{L}}^{\ast
}(\mathbf{r}+\boldsymbol{\rho }\left( t\right) )v_{i}(\mathbf{r}+\mathbf{r}%
_{1}\left( t\right) )+  \label{aux} \\
\int\limits_{-\infty }^{\infty }\mathrm{d}^{2}r~\alpha _{\mathrm{L}}(\mathbf{%
r}+\boldsymbol{\rho }\left( t\right) )v_{i}^{+}(\mathbf{r}+\mathbf{r}%
_{1}\left( t\right) ),
\end{multline}%
and $N$ and $S_{ij}(\omega )$ are given by Eqs. (\ref{LOFintens}) and (\ref%
{ModalSpectrum}), respectively.

The projections of the eigenvectors $\mathbf{v}_{i}$ onto the LOF are given
by $\left\langle \boldsymbol{\alpha}_{\mathrm{L}}|\mathbf{v}%
_{i}\right\rangle $, where the scalar product (\ref{ScalarProduct}) is used,
and the modal correlation terms $S_{ij}(\omega)$ correspond to Eq.(\ref%
{ModalSpectrum}).

Up to this point we have treated with a complete detection of the beam, that
is, we have considered an arbitrarily large detector covering the whole
transverse profile of the outgoing field. We can wonder now on the effect of
the detector size when it is finite, which corresponds to a more realistic
description.

Thus we consider now a movable detector with finite transverse size $\Sigma $%
, which allows to sweep the transverse profile of the outgoing field and
study the spatial distribution of squeezing. Mathematically, the use of a
finite size detector corresponds to limit the spatial integrations in Eq. (%
\ref{Specsum}) to a domain $\Sigma $ around the "detector position" $\mathbf{%
r}_{0}$,\ so replacing $\left\langle \boldsymbol{\alpha }_{\mathrm{L}}|%
\mathbf{v}_{i}\right\rangle $, Eq. (\ref{aux}), by%
\begin{multline}
\left\langle \boldsymbol{\alpha }_{\mathrm{L}}|\mathbf{v}_{i}\right\rangle
_{\left\{ \mathbf{r}_{0},\Sigma \right\} }\equiv  \label{PF} \\
\int\limits_{\left\{ \mathbf{r}_{0},\Sigma \right\} }\mathrm{d}^{2}r~\alpha
_{\mathrm{L}}^{\ast }\left( \mathbf{r}+\boldsymbol{\rho }\left( t\right)
\right) v_{i}\left( \mathbf{r}+\mathbf{r}_{1}\left( t\right) \right) + \\
\int\limits_{\left\{ \mathbf{r}_{0},\Sigma \right\} }\mathrm{d}^{2}r~\alpha
_{\mathrm{L}}\left( \mathbf{r}+\boldsymbol{\rho }\left( t\right) \right)
v_{i}^{+}\left( \mathbf{r}+\mathbf{r}_{1}\left( t\right) \right) ,
\end{multline}%
where $\left\{ \mathbf{r}_{0},\Sigma \right\} $ represents the spatial
region occupied by the detector, and $N$, Eq. (\ref{LOFintens}), by 
\begin{equation}
N_{\left\{ \mathbf{r}_{0},\Sigma \right\} }\equiv \int\limits_{\left\{ 
\mathbf{r}_{0},\Sigma \right\} }\mathrm{d}^{2}r\left\vert \alpha _{\mathrm{L}%
}(\mathbf{r})\right\vert ^{2}.  \label{NFinite}
\end{equation}%
Finally we can compute the squeezing spectrum measured when the finite
detector is placed at $\mathbf{r}_{0}$ through%
\begin{multline}
S_{\mathrm{out}}(\omega ;\mathbf{r}_{0})=\frac{2\gamma _{1}}{N_{\left\{ 
\mathbf{r}_{0},\Sigma \right\} }}\times  \label{spatialS} \\
\sum\limits_{i,j}\left\langle \boldsymbol{\alpha }_{\mathrm{L}}|\mathbf{v}%
_{i}\right\rangle _{\left\{ \mathbf{r}_{0},\Sigma \right\} }\left\langle 
\boldsymbol{\alpha }_{\mathrm{L}}|\mathbf{v}_{j}\right\rangle _{\left\{ 
\mathbf{r}_{0},\Sigma \right\} }S_{ij}(\omega ).
\end{multline}%
As can be noted in Eq.(\ref{spatialS}), the obtained level of squeezing
depends both on the area and position of the detector.

\section{Intensity fluctuations spectrum}

Now we apply the technique used for calculating the squeezing spectrum to
the derivation of spectrum of the intensity fluctuations. For the sake of
simplicity here we ignore the diffusive movement of the DS, i.e., we assume
that the detector can follow such motion.

The intensity fluctuations spectrum can be directly observed with a single
photodetector and is given by \cite{Glauber63}%
\begin{multline}
V_{I}\left( \omega \right) =\int\limits_{-\infty }^{\infty }\mathrm{d}%
t~e^{-i\omega t}\diint \mathrm{d}^{2}r\ \mathrm{d}^{2}r^{\prime }\times
\label{intensityspec} \\
\left\langle \delta \hat{I}_{\mathrm{out}}\left( \mathbf{r},t\right) \delta 
\hat{I}_{\mathrm{out}}\left( \mathbf{r}^{\prime },0\right) \right\rangle ,
\end{multline}%
where 
\begin{equation}
\delta \hat{I}_{\mathrm{out}}\left( \mathbf{r},t\right) =\hat{I}_{\mathrm{out%
}}\left( \mathbf{r},t\right) -\left\langle \hat{I}_{\mathrm{out}}\left( 
\mathbf{r},t\right) \right\rangle ,
\end{equation}%
and the term proportional to the intensity of the outgoing field is 
\begin{equation}
\hat{I}_{\mathrm{out}}\left( \mathbf{r},t\right) =\hat{A}_{\mathrm{out}%
}^{\dag }\left( \mathbf{r},t\right) \hat{A}_{\mathrm{out}}\left( \mathbf{r}%
,t\right) .
\end{equation}%
By making use of (\ref{commutationrel})\ and taking account of the
input-output relation (\ref{in-out}) one obtains \cite{Gatti95}%
\begin{equation}
V_{I}(\omega )=S_{\mathrm{SN}}\left[ 1+S_{I}(\omega )\right] ,
\end{equation}%
where the term corresponding to the shot noise reads 
\begin{equation}
S_{\mathrm{SN}}=2\gamma _{1}\int \mathrm{d}^{2}r~\left\langle \hat{I}%
_{1}\left( \mathbf{r},t\right) \right\rangle ,
\end{equation}%
and 
\begin{multline}
S_{I}(\omega )=\frac{4\gamma _{1}^{2}}{S_{\mathrm{SN}}}\int\limits_{-\infty
}^{\infty }\mathrm{d}t~e^{-i\omega t}\diint \mathrm{d}^{2}r\ \mathrm{d}%
^{2}r^{\prime }\times  \label{intensityS} \\
\left\langle :\delta \hat{I}_{1}\left( \mathbf{r},t\right) \delta \hat{I}%
_{1}\left( \mathbf{r}^{\prime },0\right) :\right\rangle ,
\end{multline}%
where 
\begin{subequations}
\begin{align}
\delta \hat{I}_{1}(\mathbf{r},t)& =\hat{I}_{1}(\mathbf{r},t)-\left\langle 
\hat{I}_{1}(\mathbf{r},t)\right\rangle , \\
\hat{I}_{1}(\mathbf{r},t)& =\hat{A}_{1}^{\dag }(\mathbf{r},t)\hat{A}_{1}(%
\mathbf{r},t).
\end{align}

As stated, our interest is centered on the quantum fluctuations properties
of the DS given by (\ref{ClassicalSolution}). When such form of the field is
considered, the expression of the intensity fluctuations spectrum, leading
to first order in fluctuations, reads\ 
\end{subequations}
\begin{equation}
S_{I}(\omega)=\frac{2\gamma_{1}}{\bar{N}}\int\limits_{-\infty}^{\infty }%
\mathrm{d}\tau e^{-i\omega\tau}\langle\delta\mathcal{E}_{\mathrm{I}}\left(
t+\tau\right) \delta\mathcal{E}_{\mathrm{I}}\left( t\right) \rangle,
\label{intensitySbis}
\end{equation}
where 
\begin{subequations}
\begin{align}
\delta\mathcal{E}_{\mathrm{I}}\left( t\right) & =\langle\mathbf{\bar{A}}%
_{1}\left( \mathbf{r,}t\right) \mid\mathbf{a}_{1}\left( \mathbf{r},t\right)
\rangle, \\
\bar{N} & =\int\mathrm{d}^{2}r~\left\langle \left\vert \mathcal{\bar{A}}_{1}(%
\mathbf{r})\right\vert ^{2}\right\rangle , \\
\mathbf{\bar{A}}_{1}\left( \mathbf{r}\right) & \equiv(\mathcal{\bar{A}}_{1}(%
\mathbf{r}),\mathcal{\bar{A}}_{1}^{\ast}(\mathbf{r}))^{T}.
\end{align}
Notice that $\mathbf{\bar{A}}_{1}\left( \mathbf{r}\right) $ is the vector
which corresponds to the classical DS, given by Eq.(\ref{ClassicalSolution}).

As can be noted, the intensity fluctuations spectrum, Eq.(\ref{intensitySbis}%
), has the same expression as the squeezing spectrum, Eq.(\ref{Sout}), but
with the classical DS solution acting as LOF \cite{Gatti95}. So, as it
occurred previously with the output squeezing spectrum, the intensity
fluctuations spectrum can be written in terms of the modal correlation
spectrum (\ref{ModalSpectrum})

\end{subequations}
\begin{equation}
S_{I}(\omega)=\frac{2\gamma_{1}}{\bar{N}}\sum\limits_{i,j}\left\langle 
\mathbf{\bar{A}}_{1}|\mathbf{v}_{i}\right\rangle \left\langle \mathbf{\bar{A}%
}_{1}|\mathbf{v}_{j}\right\rangle S_{ij}(\omega),  \label{intensityProject}
\end{equation}
where%
\begin{equation}
\left\langle \mathbf{\bar{A}}_{1}|\mathbf{v}_{i}\right\rangle =\int
\limits_{-\infty}^{\infty}\mathrm{d}^{2}r~\left[ \mathcal{\bar{A}}_{1}^{\ast
}(\mathbf{r})v_{i}(\mathbf{r})+\mathcal{\bar{A}}_{1}(\mathbf{r})v_{i}^{+}(%
\mathbf{r})\right] .
\end{equation}

Finally, and analogously to what we did with the squeezing spectrum, when a
finite detector of transverse size $\Sigma $ positioned at $\mathbf{r}_{0}$
is considered the intensity fluctuations are given by Eq.(\ref%
{intensityProject}) after replacing $\left\langle \mathbf{\bar{A}}_{1}|%
\mathbf{v}_{i}\right\rangle \,\ $by $\left\langle \mathbf{\bar{A}}_{1}|%
\mathbf{v}_{i}\right\rangle _{\left\{ \mathbf{r}_{0},\Sigma \right\} }$ and $%
\bar{N}$ by $\bar{N}_{\left\{ \mathbf{r}_{0},\Sigma \right\} }$ so that the
intensity fluctuations spectrum measured with a finite detector of size $%
\Sigma $ placed at $\mathbf{r}_{0}$ reads%
\begin{multline}
S_{I}(\omega ;\mathbf{r}_{0})=\frac{2\gamma _{1}}{\bar{N}_{\left\{ \mathbf{r}%
_{0},\Sigma \right\} }}\times \\
\sum\limits_{i,j}\left\langle \mathbf{\bar{A}}_{1}|\mathbf{v}%
_{i}\right\rangle _{\left\{ \mathbf{r}_{0},\Sigma \right\} }\left\langle 
\mathbf{\bar{A}}_{1}|\mathbf{v}_{j}\right\rangle _{\left\{ \mathbf{r}%
_{0},\Sigma \right\} }S_{ij}(\omega ),
\end{multline}%
where $\left\langle \mathbf{\bar{A}}_{1}|\mathbf{v}_{i}\right\rangle
_{\left\{ \mathbf{r}_{0},\Sigma \right\} }$ and $\bar{N}_{\left\{ \mathbf{r}%
_{0},\Sigma \right\} }$ are given, respectively, by Eqs.(\ref{PF}) and (\ref%
{NFinite}) when $\boldsymbol{\alpha }_{\mathrm{L}}$ is replaced by $\mathbf{%
\bar{A}}_{1}$.

\section{A general result on the squeezing of dissipative structures}

Let us assume that we can set $\mathbf{\rho}=0$ in Eq. (\ref{aux}). This
means that the LOF can be shifted in such a way that it exactly follows the
diffusive movement of the dissipative structure whose squeezing is being to
be measured. Further, let us choose the LOF $\boldsymbol{\alpha}_{\mathrm{L}%
}=\mathbf{w}_{2x}$, Eq. (\ref{lofmagic}), i.e., a LOF whose shape is $%
\alpha_{\mathrm{L}}=iG_{x}$. (We notice that the result that follows is
valid for any $\mathbf{\alpha}_{\mathrm{L}}$ that corresponds to a linear
combination of $\mathbf{w}_{2x}$ and $\mathbf{w}_{2y}$.)

By doing this it turns out that $\delta \mathcal{E}_{\mathrm{H}}\left(
t\right) =c_{2x}\left( t\right) $, see Eqs. (\ref{expansion}) and (\ref%
{deltaEh}). Standard techniques \cite{Gardiner00} applied to Eq. (\ref%
{CoefficientEvolution}) for $i=2x$%
\begin{equation}
\dot{c}_{2x}=-2\gamma _{1}c_{2x}+\sqrt{\gamma _{1}}\xi _{2x},  \label{c2x}
\end{equation}%
where $\xi _{2x}\left( t\right) =\left\langle \mathbf{w}_{2x}|\mathbf{h}%
\right\rangle $ is the noise source and $c_{2x}\left( t\right) =\left\langle 
\mathbf{w}_{2x}|\mathbf{a}_{1}\right\rangle $, allow the computation of the
stochastic correlation $\langle \delta \mathcal{E}_{\mathrm{H}}\left( t+\tau
\right) \delta \mathcal{E}_{\mathrm{H}}\left( t\right) \rangle $, that turns
out to be 
\begin{equation}
\langle \delta \mathcal{E}_{\mathrm{H}}\left( t+\tau \right) \delta \mathcal{%
E}_{\mathrm{H}}\left( t\right) \rangle =-\tfrac{1}{2}N_{\mathrm{H}%
}e^{-2\gamma _{1}\left\vert \tau \right\vert }.
\end{equation}%
Then, by using Eq. (\ref{Sout}) we get%
\begin{equation}
S_{\mathrm{out}}\left( \omega \right) =-\frac{1}{1+\left( \omega /2\gamma
_{1}\right) ^{2}},  \label{S2}
\end{equation}%
which is the main result in \cite{EPL}. Of course the same result is
obtained by using Eqs. (\ref{dij}), (\ref{ModalSpectrum}) and (\ref%
{diffusionmatrix}).

It is to be remarked that Eq. (\ref{c2x}) is analogous to that derived in 
\cite{Gatti01} for the stationary phase of the hexagonal mode appearing in
the Kerr cavity model, and that it was later interpreted in \cite{Gomila02}
as the hexagonal pattern transverse linear momentum. Then we can say that
the above result means that the transverse linear momentum of any stationary
DS appearing in the large pump detuning limit of the DOPO\ model is
perfectly squeezed in the linearized theory. This is a reasonable result as
it is immediately related to the fact that the transverse position of the DS
is completely undetermined as it diffuses with time.

Eq. (\ref{S2}) implies that $S\left( \omega=0\right) =-1$ what means that
within the validity domain of the linear theory we are using, any stationary
dissipative structure sustained by the DOPO in the large pump detuning limit
displays \textit{perfect squeezing} at $\omega=0$ when probed with the
appropriate LOF. As Eq. (\ref{S2}) is independent of the kind of dissipative
structure and of the system parameters, it is to be remarked that the result
is universal and independent of the existence of bifurcations. Let us
emphasize that the appropriate LOF ($\alpha_{\mathrm{L}}=iG_{x\left(
y\right) }$) is, in principle, easily implementable as it corresponds to the 
$\pi/2$ phase-shifted gradient of the corresponding DS envelope which can be
easily synthesized by, e.g., Fourier filtering.

In \cite{EPL} we discussed up to what extent the assumption $\boldsymbol{%
\rho }=0$ is reasonable: It is, indeed, as the diffusion of the dissipative
structures is very slow because of the large number of photons they carry,
which acts as an inertial mass \cite{EPL}.

We find it important to make here a general comment on the mathematical
technique we have presented in the previous sections. We must note that the
linearized approach we have presented is valid, in principle, when all
eigenvalues are negative: In this case all fluctuations are damped and it is
reasonable to assume that they will remain small enough. Remarkably, in our
case there exists always two null eigenvalues, which are associated with the
Goldstone modes (see Property 1 in Subsection II C). Nevertheless these null
eigenvalues do not make the linear approach invalid as the undamped
fluctuations do not concern any particular field mode but the position of
the dissipative structure, which is decoupled from the rest of fluctuations
and undergoes a continuous diffusion (as it occurs, e.g., with the phase
difference in \cite{Reynaud87,Reid88}). Numerical research carried out in
vectorial Kerr cavities \cite{Zambrini00} reinforce this confidence. In
resume:\ In spite of having a null eigenvalue, we can be confident that the
linearized description of quantum fluctuations will be reasonably accurate,
and that a nonlinear treatment \cite{Drummond05} will not lead to
dramatically different results.

\section{Squeezing properties of the DOPO bright cavity soliton}

In this second part of the article, we study in detail the squeezing
properties of the DOPO\ bright cavity soliton in the large pump detuning
limit. First, in Subsection VI.A we review the main properties of this
solution and then, in Subsection VI.B, we apply the theory developed in the
first part to it. For the sake of simplicity we shall consider in this
second part only the one--dimensional case, that is, we assume that the DOPO
works embedded in a waveguide that avoids diffraction in the $y$ transverse
dimension whilst in the $x$ transverse dimension the system aperture is
arbitrarily large. Moreover, we shall assume from now on that the LOF used
in the homodyne detection scheme can be moved, in the transverse dimension,
in such a way that its movement exactly matches the diffusive motion of the
bright cavity soliton, and thus we shall not take into account the diffusive
motion of the DS (see \cite{EPL} for a quantitative discussion in which we
show that this is a reasonable assumption).

\subsection{The DOPO bright cavity soliton}

As stated in Subsection II.A, in the limit of large pump detuning we are
considering here, the classical description of the DOPO is the PDNLSE, (Eq. (%
\ref{AdiabaticLangevinClassic}) in Appendix B). It is well known that the
one--dimensional PDNLSE supports two different types of localized structures
(cavity solitons in our context): Dark cavity solitons (tanh--type localized
structures) in the self-defocusing case $\sigma =-1$, and bright\ cavity
solitons (sech--type localized structures) in the self-focusing case $\sigma
=+1$ \cite{Fauve90,Miles84,Elphick89,Barashenkov91}. We shall treat in the
following the bright cavity soliton and thus we take $\sigma =+1$.

Before going on we find it convenient to review the main solutions of the
PDNLSE. This equation has only two free parameters (the parametric pump
parameter $\mu $ and the cavity detuning $\Delta _{1}$) as the rest of
parameters ($\gamma _{1}$, $l_{1}$, and $\kappa ^{2}$) can be easily removed
by normalizing the time and space coordinates as well as the field amplitude
(see Eq. (\ref{PDNLSE}) in Appendix B). When $\Delta _{1}<0$, the trivial
state, $\mathcal{\bar{A}}_{1}(x)=0$, undergoes a supercritical bifurcation
towards a patterned state (a roll, or stripe, pattern) at $\mu =1$.
Contrarily, when $\Delta _{1}>0$ the bifurcation affecting the trivial state
is subcritical and occurs at $\mu =\mu _{0}\equiv \sqrt{1+\Delta _{1}^{2}}$.
For $\mu >\mu _{0}$, and positive $\Delta _{1}$, dynamic patterns are found 
\cite{Longhi95}.

The bright cavity soliton has the explicit expression%
\begin{equation}
\mathcal{\bar{A}}_{1,\mathrm{BS}}(x)=\sqrt{2}\beta ~e^{i\phi }\func{sech}%
(\beta x),  \label{BS}
\end{equation}%
with 
\begin{subequations}
\label{solitonsPGLE}
\begin{align}
\beta ^{2}& =\Delta _{1}\pm \sqrt{\mu ^{2}-1},  \label{beta} \\
\ \cos (2\phi )& =\mu ^{-1}.
\end{align}%
This solution exists for $\Delta _{1}>0$ and $1<\mu <\mu _{0}$ \cite%
{Alexeeva99}, and is stable in a wide domain of parameters, although for
large enough $\Delta _{1}$ it becomes unstable, through a Hopf bifurcation,
thus appearing temporal oscillations (self-pulsing solitons) \cite{Bondila95}%
. Then the bright cavity soliton can undergo three different bifurcations:
(i) a tangent bifurcation at $\mu =1$ (the BS does not exist for $\mu <1$),
(ii) a bifurcation at $\mu =\mu _{0}$ (the BS does not exist for $\mu >\mu
_{0}$), and (iii) a Hopf bifurcation for large enough $\Delta _{1}$ that
transforms the stationary BS into a self--pulsing BS. In Fig. 1 we represent
in the $\left\langle \Delta _{1},\mu \right\rangle $ plane the domains of
existence of the different solutions we have just commented.

It is convenient to briefly comment here on the similarities between the
bright cavity soliton and the temporal soliton of the nonlinear Schr\"{o}%
dinger equation (NLSE). The NLSE is a conservative equation that can be
written without free parameters. Then, there is a family of sech--type
solitons, that coexist, and for which certain quantities (such as the
energy) are conserved. This is very different from the PDNLSE which is a
nonconservative equation governed by two parameters. For given $\mu$ and $%
\Delta_{1}$, the sech--type solution is unique and can be stable or
unstable. In fact the bright cavity soliton is not a true soliton and its
name can be misleading when comparisons between these similar but actually
quite different equations (NLSE and PDNLSE) are made. In particular, these
differences manifest in the spectra of $\mathcal{L}$ and$~\mathcal{L}^{\dag}$%
.

We have shown that in order to calculate the squeezing spectrum, given by
Eq. (\ref{Specsum}), one needs to evaluate the spectra (\ref{Eigensystem})
of both $\mathcal{L}$ and$~\mathcal{L}^{\dag }$ (\ref{LinearOperator}) for $%
\mathcal{\bar{A}}_{1}(\mathbf{r})=\mathcal{\bar{A}}_{1,\mathrm{BS}}(x)$, Eq.
(\ref{BS}). The spectrum of $\mathcal{L}$ in this case has been extensively
studied \cite{Barashenkov91}. It consists of a continuous spectrum, with
eigenvalues of the form 
\end{subequations}
\begin{subequations}
\begin{align}
\lambda _{s}(k)& =-1+s\sqrt{\mu ^{2}-\Delta _{k}^{2}}, \\
\Delta _{k}& =\Delta _{1}+sk^{2},\text{\ \ }s=\pm 1,\ \ \ k\in \lbrack
0,\infty )
\end{align}%
and of a discrete spectrum with eigenvalues $\{\lambda _{i}\}_{i=1}^{D}$.
The spectra of $\mathcal{L}$ and$~\mathcal{L}^{\dag }$ need to be computed
numerically, as analytical expressions can not be derived in general, which
we have done by adapting the Fourier method described in \cite{Alexeeva99}.
Nevertheless, some eigenvectors of the discrete spectrum can be computed
analytically for some parameter sets. In particular, we have derived the
four eigenvectors corresponding to the bifurcation occurring at $\mu =1$.
These eigenvectors are explicitly given in Appendix B.

A crucial point is whether the set of eigenvectors of $\mathcal{L}$ and$~%
\mathcal{L}^{\dag}$ form a (biorthonormal) basis or not. We commented above
that for the similar problem of the nonlinear Schr\"{o}dinger equation,
describing temporal fiber solitons, the set of eigenvectors of $\mathcal{L}$
and$~\mathcal{L}^{\dag}$ do not form a base \cite{Kozlov03}. As in our case
analytical expressions are not available, our strategy has consisted in
discretizing the transverse spatial coordinate and numerically diagonalize $%
\mathcal{L}$ and$~\mathcal{L}^{\dag}$. Our numerical results show that the
number of independent eigenvectors coincides with the dimension of the
matrices, what constitutes a numerical proof that for this particular
transverse pattern, the eigenvectors of $\mathcal{L}$ and$~\mathcal{L}%
^{\dag} $ form a (biorthonormal) basis, whereas this does not happen in the
conservative case. We see that the NLSE and the PDNLSE are very different
problems indeed.

\subsection{Squeezing properties of the unidimensional bright cavity soliton}

Now we are in conditions for studying the squeezing properties of the DOPO\
bright cavity soliton. Of course the amount of squeezing to be obtained will
depend on the local oscillator field (LOF) one uses in the homodyne
measurements and, in general, on the parameter values (pump and detuning).
We shall proceed in our study as follows: First, in Subsection VI.B.1, we
shall particularize the general result obtained in \cite{EPL} and derived
above in Section III for the bright cavity soliton. Then, in Subsection
VI.B.2, we shall consider the squeezing at the bifurcation points (which
have been discussed in Subection VI.A). Finally, in Subsection VI.B.3 we
shall consider the case of a plane--wave LOF. In this last case we discuss
first how the amount of squeezing changes with the parameters (for
arbitrarily large detectors), and then how the level of squeezing depends on
the size of the photodetectors when these have a finite transverse dimension.

\subsubsection{Squeezing of the cavity soliton linear momentum}

In Section VI we showed that there is a special mode that is perfectly
squeezed in the linear approximation, which is equivalent to saying that the
transverse linear momentum of the DS is perfectly squeezed. We saw that the
fluctuations of this mode are detected by choosing the LOF $\boldsymbol{%
\alpha }_{\mathrm{L}}(x)=\mathbf{w}_{2x}(x)$, that is, $\alpha _{\mathrm{L}}(%
\mathbf{r})=i\partial _{x}\mathcal{\bar{A}}_{1,\mathrm{BS}}\left( x\right) $%
, see Eq.(\ref{lofmagic}). In Fig. 2 we represent the amplitude of this LOF
with a dashed line together with the soliton amplitude, given by Eq. (\ref%
{BS}). (As the phase factor $\exp \left( i\phi \right) $ is space
independent, Eq. (\ref{BS}), it has been discarded in making this plot.)

The similarity between $w_{2x}(x)$ and the Gauss--Hermite function 
\end{subequations}
\begin{equation}
GH_{1}(x)\equiv ie^{i\phi }xe^{-\frac{1}{2}(x/\xi )^{2}},  \label{GH}
\end{equation}%
is immediate (in Fig. 2 the Gauss--Hermite function is also plotted), what
suggests the use of this function, which is relatively easy to generate, as
a LOF. In order to calculate the level of squeezing that would be detected
by using the Gauss--Hermite LOF, Eq. (\ref{GH}), we take advantage of the
general theory presented in the first part of this article:\ We expand the
LOF on the basis of $\mathcal{L}^{\dag }$ as%
\begin{equation}
\boldsymbol{\alpha }_{\mathrm{L}}=\sum_{i}\left\langle \mathbf{v}_{i}|%
\boldsymbol{\alpha }_{\mathrm{L}}\right\rangle \mathbf{w}_{i},
\end{equation}%
excluding the Goldstone modes, and the squeezing spectrum is given by%
\begin{equation}
S_{\mathrm{out}}(\omega )=\frac{2\gamma }{N}\sum\limits_{i,j}\left\langle 
\mathbf{v}_{i}|\boldsymbol{\alpha }_{\mathrm{L}}\right\rangle \left\langle 
\mathbf{v}_{j}|\boldsymbol{\alpha }_{\mathrm{L}}\right\rangle S_{ij}(\omega
),
\end{equation}%
where the modal correlation matrix, Eq.(\ref{ModalSpectrum}), is evaluated
from the computed spectra of $\mathcal{L}$ and$~\mathcal{L}^{\dag }$. The
accuracy of the numerical method was checked by computing $S_{\mathrm{out}%
}(\omega )$ when $\boldsymbol{\alpha }_{\mathrm{L}}(x)=\mathbf{w}_{2}(x)$,
yielding an error less than $10^{-13}$. The influence of the Gauss--Hermite
LOF width and position was already presented in \cite{EPL}, and we reproduce
it in Fig. 3 for the sake of completitude. Notice that the level of
squeezing is quite large even when the LOF position or width are not well
matched to those of mode $\mathbf{w}_{2}(x)$.

\subsubsection{Squeezing at the bifurcation points}

At the bifurcations points there is, at least, one null eigenvalue, apart
from that corresponding to the Goldstone mode. This implies the existence of
a mode, different to $\mathbf{w}_{2}(x)$, whose eigenvalue reaches its
minimum value. This mode is expected to be perfectly squeezed in the linear
approximation. The squeezing of these modes is the equivalent to the usual
squeezing at the bifurcation points that has been repeatedly studied in a
large number of nonlinear cavities \cite{Meystre91,Walls94}.

We have seen that the bright cavity soliton can undergo three different
bifurcations at $\mu=1$, $\mu=\mu_{0}$, and $\mu=\mu_{\mathrm{HB}}$, see
Fig. 1. Let us consider first the bifurcation occurring at $\mu=1$ (the
analytic expression of the mode that has $\lambda=-2$ is given in the
Appendix B, mode $\mathbf{w}_{3}(x)$). Then, by taking $\boldsymbol{\alpha}_{%
\mathrm{L}}(x)=\mathbf{w}_{3}(x)$, one obtains the squeezing spectra that we
have represented in Fig. 4 for $\Delta_{1}=1.2$ and three values of $\mu$.
Notice that $S_{\mathrm{out}}\left( \omega=0\right) =-1$ for $\mu=1$ (full
line); for $\mu=1.0001$ (dashed line) the maximum squeezing does not reach $%
-1$. Furthermore, as $\mu$ departs from unity the maximum squeezing does not
occur at $\omega=0$, but at a slightly different frequency. For $\mu=1.01$,
the maximum level of squeezing is close to $-0.75$. That is, the squeezing
degrades quickly as the system departs from this bifurcation. A similar
behavior is exhibited at the bifurcation occurring at $\mu=\mu_{0}=\sqrt{%
1+\Delta_{1}^{2}}$ and we shall not enter into quantitative details.

Let us finally consider the Hopf bifurcation. For $\mu =\mu _{\mathrm{HB}}$
there are two eigenmodes, let us denote them as $\mathbf{w}_{\mathrm{HB}+}$
and $\mathbf{w}_{\mathrm{HB}-}$, for which the eigenvalues read $\lambda
_{\pm }=-2$ $\pm i\omega _{\mathrm{HB}}$ ($\omega _{\mathrm{HB}}$ is the
oscillation frequency at the Hopf bifurcation), i.e., there are two most
damped eigenmodes which are expected to exhibit maximum squeezing. But in
the Hopf bifurcation there cannot be a LOF that matches these squeezed
modes: Remind that the LOF vector was defined as $\boldsymbol{\alpha }_{%
\mathrm{L}}=\left( \alpha _{\mathrm{L}},\alpha _{\mathrm{L}}^{\ast }\right)
^{T}$, and at the Hopf bifurcation the eigenvectors with Re $\lambda =-2$, $%
\mathbf{w}_{\mathrm{HB}\pm }=\left( w_{\mathrm{HB}\pm },w_{\mathrm{HB}\pm
}^{+}\right) ^{T}$, do not verify $w_{\mathrm{HB}\pm }^{+}=w_{\mathrm{HB}\pm
}^{\ast }$. This can be appreciated in Fig. 5 where we have represented the
imaginary parts of $w_{\mathrm{HB}+}^{\ast }$ and $w_{\mathrm{HB}+}^{+}$
(also the real parts, not shown, are quite different). This fact makes it
impossible to find a LOF such that $\left( \boldsymbol{\alpha }_{\mathrm{L}}|%
\mathbf{v}_{\mathrm{HB}}\right) =1$. Then perfect noise reduction is never
achieved at the Hopf bifurcation.

The amount of squeezing attainable at the Hopf bifurcation is represented in
Fig. 6 where the squeezing spectrum has been represented for two choices of
the LOF: $\boldsymbol{\alpha }_{\mathrm{L}}(x)=\mathbf{w}_{\mathrm{HB}+}(x)$%
, in dashed line, and $\boldsymbol{\alpha }_{\mathrm{L}}(x)=\mathbf{w}_{%
\mathrm{HB}+}(x)+\mathbf{w}_{\mathrm{HB}-}(x)$, in full line. Interestingly
the maximum squeezing is larger for the latter choice. Notice also that, in
contrast to the other bifurcations and as it is well known, at the Hopf
bifurcation the maximum level of squeezing is reached at a frequency
different from $\omega =0$, which corresponds to the self-pulsing frequency
of the bright cavity soliton.

\subsubsection{Squeezing with a plane-wave local oscillator}

We consider now the particular case of the squeezing properties of the
bright cavity soliton when measured in a homodyning experiment, probed with
a plane-wave LOF, which is of obvious relevant interest as this is the
simplest LOF.

At a first stage we deal, as in the preceding subsections, with a complete
beam detection, that is, the detector is assumed to completely cover the
transverse extension of the outgoing quantum cavity soliton. Fig. 7 shows
the squeezing spectra obtained for five different values of the pump $\mu $
and $\Delta _{1}=1.2$. The maximum of squeezing is achieved at frequency $%
\omega =0$ irrespective of the pump value (which is obviously limited to $1<$
$\mu <$ $\mu _{0}=\sqrt{1+\Delta _{1}^{2}}$, see Fig. 1). A high degree of
squeezing, sustained for a large range of parameters setting, is obtained.
This is more clearly seen in the inset of Fig. 7, where $S(\omega =0)$ is
depicted as a function of pump for $\Delta _{1}=1.2$.

We have also calculated the influence of the detuning: In Fig. 8 we plot the
same as in Fig. 7 but for fixed $\mu =1.2$ and as a function of detuning $%
\Delta _{1}$. Again large squeezing levels are obtained in all the domain of
existence of the cavity soliton (for detunings larger than those represented
in Fig. 8, the cavity soliton becomes Hopf unstable as already commented).

We focus now on the homodyning detection when using finite size detectors.
When a detector, with size $\Sigma $ and positioned at $x=x_{0}$ is used,
the squeezing spectrum detected is given by Eq. (\ref{spatialS}),
particularized to the BS solution. In the particular case of a plane-wave
LOF, the calculations are considerably simplified, as $\mathbf{\alpha }_{L}$
is constant. We note, as well, that the phase of the plane-wave LOF is a
free parameter, which is allowed to vary in order to obtain the maximum
level of squeezing.

In Fig. 9, we represent the spatial distribution of squeezing, for $\mu
=\Delta _{1}=1.2$ and at $\omega =0$, when the finite size detector is
displaced across the transverse dimension. We consider three different
values of the detector size $\Sigma =\Delta x/\beta $ normalized to the BS
width $\beta $, see Eq. (\ref{beta}), as indicated in the figure. In the
three cases the phase of the plane-wave LOF (not shown) has been chosen in
order to obtain the maximum squeezing at each position of the detector in
the transverse plane. A clear conclusion can be extracted from these plots:
The smaller is the detector size, the smaller is the squeezing level one
attains. This is to be expected as the smaller the photodetector is, the
larger influence of high spatial frequency modes, i.e., with the smaller
photodetector the influence of all sort of nonsqueezed modes is larger than
with the larger photodetectors which filter out high frequency modes. One
more conclusion one can extract from the figure is that the squeezing level
is larger, for all detector sizes, at the center of the soliton ($x=0$),
i.e., the soliton is more squeezed than the vacuum (for large $x$ the
soliton amplitude tends to zero and the squeezing is due to the squeezed
vaccum).

Finally we focus on the influence of the detector size on the degree of
squeezing reached when the detector is placed at the exact center of the
soliton. Fig. 10 shows the maximum squeezing reached in these conditions at
frequency $\omega =0$ when a detector of normalized size $\Sigma $ is
considered. Several features can be appreciated in the figure. As a general
trend, the squeezing level degrades as the detector size is reduced, i.e.,
the maximum level of squeezing is found for a very large detector. But,
interestingly, for $1.5<\Sigma <3$, the squeezing decreases with the
increase of the detector size. Obviously, for detectors whose size falls in
this region what is happening is that the fluctuations being detected come
from both the BS and the trivial solution (below threshold emission)
existing far from the detector. In fact we see that the phase of the local
oscillator that optimizes the squeezing level (represented with a dashed
line, left vertical axis), which is almost insensitive to the detector size
for $\Sigma <1$, changes rapidly in this detector width region. Then, for
large detectors, $\Sigma >10$, the monotonic increase of squeezing with the
detector width is recovered and the optimum phase for the LOF\ is again
almost insensitive to the detector size.

\section{Conclusions}

In this article we have developed a general theory for the analysis of
linearized quantum fluctuations of optical dissipative structures generated
in wide aperture nonlinear optical cavities. Although we have done this
explicitly for the special case of the degenerate optical parametric
oscillator in the large pump detuning limit, our method can be easily
generalized to any other nonlinear cavity. The method consists, in short, in
expanding the fluctuations in the biorthonormal base that the eigenvectors
of the linear deterministic operator of the linearized Langevin equations of
the system. This technique allows, in particular, the identification of a
special mode which is perfectly squeezed (in the linear approximation). The
perfect squeezing occurs irrespective of the nonlinear cavity parameter
values, and the special mode can be identified with the transverse linear
momentum of the dissipative structure that is being emitted. It must be
emphasized that the existence of this squeezed mode is a genuine transverse
effect, i.e., it is associated to the symmetry breaking introduced by the
existence of dissipative structures.

Then we have applied this theory to the study of squeezing of a particular
dissipative structure, namely, the bright cavity soliton. In particular we
have analyzed the squeezing occurring at the different bifurcations that
this dissipative structure can undergo at the same time that the appropriate
LOF for detecting it in each situation. Then we have studied the squeezing
detected when a plane wave LOF is used and analyzed its dependence on the
parameter values. Finally we have considered also finite size detectors. We
have shown that for large detectors the squeezing level is large almost
independently of the system parameters. For finite size detectors, we have
analyzed the spatial distribution of squeezing.

We gratefully acknowledge fruitful discussions with A. Gatti, K. Staliunas
and J. A. de Azc\'{a}rraga. This work has been supported by the Spanish
Ministerio de Educaci\'{o}n y Ciencia and the European Union FEDER through
Projects BFM2002-04369-C04-01, FIS2005-07931-C03-01 and -02 and Programa
Juan de la Cierva.

\section{Appendix A}

In this Appendix the DOPO model used along the paper is derived. The
Hamiltonian describing the DOPO in the interaction picture is given by \cite%
{Gatti97}%
\begin{equation}
\hat{H}=\hat{H}_{\mathrm{free}}+\hat{H}_{\mathrm{int}}+\hat{H}_{\mathrm{ext}%
},
\end{equation}%
where%
\begin{equation}
\hat{H}_{\mathrm{free}}=\hbar \sum\limits_{n=0,1}\gamma _{n}\int \mathrm{d}%
^{2}r~\hat{A}_{n}^{\dag }\left( \Delta _{n}-l_{n}^{2}\nabla ^{2}\right) \hat{%
A}_{n},
\end{equation}%
governs the free evolution of the intracavity fields in the paraxial
approximation,%
\begin{equation}
\hat{H}_{\mathrm{int}}=\frac{\hbar g}{2}\int^{2}\mathrm{d}^{2}r~i\left[ \hat{%
A}_{0}\left( \hat{A}_{1}^{\dag }\right) ^{2}-\hat{A}_{0}^{\dag }\left( \hat{A%
}_{1}\right) ^{2}\right] ,
\end{equation}%
describes the nonlinear interaction, and%
\begin{equation}
\hat{H}_{\mathrm{ext}}=\hbar \int \mathrm{d}^{2}r~i\left[ \mathcal{E}_{%
\mathrm{int}}\hat{A}_{0}^{\dag }-\mathcal{E}_{\mathrm{int}}^{\ast }\hat{A}%
_{0}\right] ,
\end{equation}%
accounts for the coherent driving. In the above expressions $l_{n}=c/\sqrt{%
2\omega _{n}\gamma _{n}}$ is the diffraction length for the field $\hat{A}%
_{n}$, $\Delta _{0}=\left( \omega _{0}-2\omega _{\mathrm{s}}\right) /\gamma
_{0}$ and $\Delta _{1}=\left( \omega _{1}-\omega _{\mathrm{s}}\right)
/\gamma _{1}$ are the (adimensional) pump and signal detuning parameters,
respectively, $\nabla ^{2}=\partial ^{2}/\partial x^{2}+\partial
^{2}/\partial y^{2}$ is the transverse Laplacian operator, and $g$ is the
(real) nonlinear coupling coefficient, given by 
\begin{equation}
g=\frac{3\chi ^{(2)}\omega _{\mathrm{s}}}{2(2\pi n)^{3}}\sqrt{\frac{\hbar
\omega _{\mathrm{s}}}{\varepsilon _{0}L_{z}}},  \label{defg}
\end{equation}%
with $\chi ^{(2)}$ the relevant nonlinear susceptibility matrix element, $n$
the common value of the refractive index of the crystal at pump and signal
wavelengths (a type I DOPO is considered), and $L_{z}$ the thickness of the
crystal along the resonator axis.

From the above Hamiltonian one obtains the master equation governing the
evolution of the density matrix $\hat{\rho}$ of the intracavity modes,%
\begin{equation}
\frac{\partial }{\partial t}\hat{\rho}=\frac{1}{i\hbar }\left[ \hat{H},\hat{%
\rho}\right] +\widehat{\Lambda \rho },  \label{MasterEq}
\end{equation}%
where the Liouvillian term%
\begin{multline}
\widehat{\Lambda \rho }=\sum\limits_{n=0,1}\gamma _{n}\dint \mathrm{d}^{2}r%
\left[ 2\hat{A}_{n}(\mathbf{r},t)\hat{\rho}\hat{A}_{n}^{\dag }(\mathbf{r}%
,t)-\right. \\
\left. \hat{\rho}\hat{A}_{n}^{\dag }(\mathbf{r},t)\hat{A}_{n}(\mathbf{r},t)-%
\hat{A}_{n}(\mathbf{r},t)\hat{A}_{n}^{\dag }(\mathbf{r},t)\hat{\rho}\right] ,
\end{multline}%
models the coupling between the system and the external reservoir through
the output mirror.

Passing to the the generalized $P$ representation \cite{Drummond80} one can
transform the master equation (\ref{MasterEq}) into an equivalent
Fokker-Planck equation for a quasiprobability density (denoted by $P$),
following standard methods (see, e.g. \cite{Gatti97}), the result being:

\begin{multline}
\frac{\partial }{\partial t}P(\mathbf{A})=\left\{ \dint \mathrm{d}^{2}r~%
{\LARGE [}\frac{\partial }{\partial \mathcal{A}_{0}}\left( -\gamma _{0}L_{0}%
\mathcal{A}_{0}+\frac{g}{2}\mathcal{A}_{1}^{2}-\mathcal{E}_{\mathrm{in}%
}\right) +\right.  \label{Fokker-Planck} \\
\frac{\partial }{\partial \mathcal{A}_{0}^{+}}\left( -\gamma _{0}L_{0}^{\ast
}\mathcal{A}_{0}^{+}+\frac{g}{2}\mathcal{A}_{1}^{+2}-\mathcal{E}_{\mathrm{in}%
}^{\ast }\right) + \\
\frac{\partial }{\partial \mathcal{A}_{1}}\left( -\gamma _{1}L_{1}\mathcal{A}%
_{1}-g\mathcal{A}_{1}^{+}\mathcal{A}_{0}\right) + \\
\frac{\partial }{\partial \mathcal{A}_{1}^{+}}\left( -\gamma _{1}L_{1}^{\ast
}\mathcal{A}_{1}^{+}-g\mathcal{A}_{1}\mathcal{A}_{0}^{+}\right) + \\
\left. \frac{g}{2}\left( \frac{\partial ^{2}}{\partial \mathcal{A}_{1}^{2}}%
\mathcal{A}_{0}+\frac{\partial ^{2}}{\partial \mathcal{A}_{1}^{+2}}\mathcal{A%
}_{0}^{+}\right) {\LARGE ]}\right\} P(\mathbf{A}).
\end{multline}%
In the above expression $\mathbf{A}=(\mathcal{A}_{0},\mathcal{A}_{0}^{+},%
\mathcal{A}_{1},\mathcal{A}_{1}^{+})$, and%
\begin{equation}
L_{j}=-(1+i\Delta _{j})+il_{j}^{2}\nabla ^{2}.  \label{L}
\end{equation}

In its turn, a Fokker-Planck equation, here Eq. (\ref{Fokker-Planck}), can
be transformed into an equivalent classical-looking set of stochastic
differential equations (so called Langevin equations) via Ito rules \cite%
{Gardiner00}. In our case they read 
\begin{subequations}
\label{completeLangevin}
\begin{align}
\frac{\partial }{\partial t}\mathcal{A}_{0}& =\gamma _{0}L_{0}\mathcal{A}%
_{0}-\frac{g}{2}\mathcal{A}_{1}^{\ \ 2}+\mathcal{E}_{\mathrm{in}}, \\
\frac{\partial }{\partial t}\mathcal{A}_{0}^{+}& =\gamma _{0}L_{0}^{\ast }%
\mathcal{A}_{0}^{+}-\frac{g}{2}\mathcal{A}_{1}^{+}{}^{2}+\mathcal{E}_{%
\mathrm{in}}^{\ast }, \\
\frac{\partial }{\partial t}\mathcal{A}_{1}& =\gamma _{1}L_{1}\mathcal{A}%
_{1}+g\mathcal{A}_{1}^{+}\mathcal{A}_{0}+\sqrt{g\mathcal{A}_{0}}\eta (%
\mathbf{r},t), \\
\frac{\partial }{\partial t}\mathcal{A}_{1}^{+}& =\gamma _{1}L_{1}^{\ast }%
\mathcal{A}_{1}^{+}+g\mathcal{A}_{1}\mathcal{A}_{0}^{+}+\sqrt{g\mathcal{A}%
_{0}^{+}}\eta ^{+}(\mathbf{r},t),
\end{align}%
with $\eta (\mathbf{r},t)$ and $\eta ^{+}(\mathbf{r},t)$ independent, real
white Gaussian noises of zero average and correlations given by 
\end{subequations}
\begin{subequations}
\label{noise}
\begin{gather}
\left\langle \eta ^{+}(\mathbf{r}^{\prime },t^{\prime }),\eta (\mathbf{r}%
,t)\right\rangle =0, \\
\left\langle \eta ^{+}(\mathbf{r},t),\eta ^{+}(\mathbf{r}^{\prime
},t^{\prime })\right\rangle =\delta (\mathbf{r}-\mathbf{r}^{\prime })\delta
(t-t^{\prime }), \\
\left\langle \eta (\mathbf{r},t),\eta (\mathbf{r}^{\prime },t^{\prime
})\right\rangle =\delta (\mathbf{r}-\mathbf{r}^{\prime })\delta (t-t^{\prime
}).
\end{gather}

Note that, if $\mathcal{A}_{i}^{+}$ is interpreted as $\mathcal{A}_{i}^{\ast
}$ and the noise terms are ignored (classical limit), Eqs. (\ref%
{completeLangevin}) coincide with the classical equations for a planar DOPO 
\cite{Oppo94}.

Equations (\ref{completeLangevin}) are already set in a convenient way to
apply the method we develop in this paper. However we shall use a simpler
form of them obtained in the limit of large pump detuning, defined by $%
\left\vert \Delta _{0}\right\vert >>1,\gamma _{0}/\gamma _{1},\left\vert
\Delta _{1}\right\vert $, which allows the adiabatic elimination of the pump
field \cite{Longhi97,Trillo97} as 
\end{subequations}
\begin{subequations}
\label{AdiabaticSol}
\begin{align}
\mathcal{A}_{0}& =\mathcal{A}_{0,\mathrm{ad}}\equiv \frac{-i}{\gamma
_{0}\Delta _{0}}\left( \mathcal{E}_{\mathrm{in}}-\frac{g}{2}\mathcal{A}%
_{1}^{2}\right) , \\
\mathcal{A}_{0}^{+}& =\mathcal{A}_{0,\mathrm{ad}}^{+}\equiv \frac{i}{\gamma
_{0}\Delta _{0}}\left( \mathcal{E}_{\mathrm{in}}^{\ast }-\frac{g}{2}\mathcal{%
A}_{1}^{+}{}^{2}\right) .
\end{align}%
We note that these expressions correct some typos appearing in the
corresponding expressions in \cite{EPL}. The remaining Langevin equations
for the signal field are obtained by substitution of the limit solution (\ref%
{AdiabaticSol})\ into (\ref{completeLangevin}). By taking, without loss of
generality, the external pump field amplitude as a purely imaginary quantity
whose imaginary part has the same sign as the pump detuning $\Delta _{0}$,
i.e. $\mathcal{E}_{\mathrm{in}}=i\sigma \left\vert \mathcal{E}_{\mathrm{in}%
}\right\vert $ with $\sigma =\func{sign}\Delta _{0}$, they become Eqs. (\ref%
{AdiabaticLangevin}) in Sec. \ref{DOPOmodel}, which is the model we shall
consider.

\section{Appendix B}

In the classical limit ($\mathcal{A}_{1}^{+}\rightarrow\mathcal{A}%
_{1}^{\ast} $, and $\eta,\eta^{+}\rightarrow0$), Eqs. (\ref%
{AdiabaticLangevin}) reduce to 
\end{subequations}
\begin{equation}
\frac{\partial}{\partial t}\mathcal{A}_{1}\left( \mathbf{r},t\right)
=\gamma_{1}\left( L_{1}\mathcal{A}_{1}+\mu\mathcal{A}_{1}^{\ast}+i\frac{%
\sigma}{\kappa^{2}}\left\vert \mathcal{A}_{1}\right\vert ^{2}\mathcal{A}%
_{1}\right) ,  \label{AdiabaticLangevinClassic}
\end{equation}
$L_{1}=-(1+i\Delta_{1})+il_{1}^{2}\nabla^{2}$, which is a version of the so
called parametrically driven, nonlinear Schr\"{o}dinger equation (PDNLSE),
which is a universal model for pattern formation in parametrically driven
systems (see, e.g., \cite{deValcarcel02} for a list of systems described by
the PDNLSE). A convenient normalization of that equation is obtained by
introducing the following dimensionless quantities: time $T=\gamma_{1}t$,
spatial coordinates $\left( X,Y\right) =\left( x/l_{1},y/l_{1}\right) $, and
field $\psi=\mathcal{A}_{1}/\kappa$, so that Eq. (\ref%
{AdiabaticLangevinClassic}) becomes%
\begin{equation}
\partial_{T}\psi=\mu\psi^{\ast}-(1+i\Delta_{1})\psi+i\left(
\partial_{X}^{2}+\partial_{Y}^{2}\right) \psi+i\sigma\left\vert
\psi\right\vert ^{2}\psi.  \label{PDNLSE}
\end{equation}

For the bright cavity soliton, Eq. (\ref{BS}), operators $\mathcal{L}$ and$~%
\mathcal{L}^{\dag}$ take the form 
\begin{subequations}
\begin{equation}
\mathcal{L}\mathcal{=}\left( 
\begin{array}{cc}
\mathcal{L}_{1} & \mathcal{\bar{A}}_{0} \\ 
\mathcal{\bar{A}}_{0}^{\ast} & \mathcal{L}_{1}^{\ast}%
\end{array}
\right) ,\ \ \mathcal{L}^{\dag}\mathcal{=}\left( 
\begin{array}{cc}
\mathcal{L}_{1}^{\ast} & \mathcal{\bar{A}}_{0} \\ 
\mathcal{\bar{A}}_{0}^{\ast} & \mathcal{L}_{1}%
\end{array}
\right) ,
\end{equation}
where 
\end{subequations}
\begin{subequations}
\begin{align}
\mathcal{L}_{1} & =-(1+i\Delta_{1})+i\nabla^{2}+i\left( \frac{2\beta }{\kappa%
}\right) ^{2}\func{sech}^{2}(\beta x), \\
\mathcal{\bar{A}}_{0} & =\mu+i\left( \frac{2\beta}{\kappa}\right)
^{2}~e^{2i\phi}\func{sech}^{2}(\beta x)
\end{align}
and 
\end{subequations}
\begin{subequations}
\begin{align}
\beta^{2} & =\Delta_{1}\pm\sqrt{\mu^{2}-1}, \\
\ \cos(2\phi) & =\mu^{-1}.
\end{align}

Then, for $\mu=1$ one can calculate the discrete eigenvalues analytically.
The result is that the discrete eigenvectors of $\mathcal{L}$ and$~\mathcal{L%
}^{\dag}$ with null eigenvalue are 
\end{subequations}
\begin{subequations}
\begin{align}
\mathbf{v}_{1} & =-\mathcal{ST}\left( 
\begin{array}{c}
e^{i\phi} \\ 
e^{-i\phi}%
\end{array}
\right) ,\ \ \  \\
\mathbf{w}_{1} & =-\mathcal{S}\left( 
\begin{array}{c}
\left( x+i\mathcal{T}\right) e^{i\phi} \\ 
\left( x-i\mathcal{T}\right) e^{-i\phi}%
\end{array}
\right) , \\
\mathbf{v}_{4} & =i\beta^{-1/2}\mathcal{S}\left( 
\begin{array}{c}
\left[ \beta^{2}+i\left( x\mathcal{T}-1\right) \right] e^{i\phi} \\ 
-\left[ \beta^{2}-i\left( x\mathcal{T}-1\right) \right] e^{-i\phi}%
\end{array}
\right) , \\
\mathbf{w}_{4} & =\beta^{1/2}\mathcal{S}\left( 
\begin{array}{c}
e^{i\phi} \\ 
e^{-i\phi}%
\end{array}
\right) ,
\end{align}
and that the discrete eigenvectors of $\mathcal{L}$ and$~\mathcal{L}^{\dag}$
with eigenvalue $\lambda=-2$ are 
\end{subequations}
\begin{subequations}
\begin{align}
\mathbf{v}_{2} & =-i\beta^{1/2}\ \mathcal{S}\left( 
\begin{array}{c}
\left( x+i\mathcal{T}\right) e^{i\phi} \\ 
-\left( x-i\mathcal{T}\right) e^{-i\phi}%
\end{array}
\right) , \\
\mathbf{w}_{2} & =-i\beta^{-1/2}\mathcal{ST}\left( 
\begin{array}{c}
e^{i\phi} \\ 
-e^{-i\phi}%
\end{array}
\right) , \\
\mathbf{v}_{3} & =i\beta\mathcal{S}\left( 
\begin{array}{c}
e^{i\phi} \\ 
-e^{-i\phi}%
\end{array}
\right) , \\
\mathbf{w}_{3} & =-\beta^{-1}\mathcal{S}\left( 
\begin{array}{c}
\left[ \beta^{2}+i\left( x\mathcal{T}-1\right) \right] e^{i\phi} \\ 
\left[ \beta^{2}-i\left( x\mathcal{T}-1\right) \right] e^{-i\phi}%
\end{array}
\right) .
\end{align}
In the above expressions $\mathbf{v}_{1}$ is the Goldstone mode, Eq. (\ref%
{goldstone}), and we have introduced the quantities 
\end{subequations}
\begin{equation}
\mathcal{S}=\sqrt{\frac{\beta}{2}}\func{sech}(\beta x),\ \ \mathcal{T}%
=\beta\tanh(\beta x).
\end{equation}
Let us finally notice that for $\mu=1$, it turns out that $%
\beta^{2}=\Delta_{1}$ and $\cos\left( 2\phi\right) =1$.

\newpage \bigskip {\large FIGURE CAPTIONS}\bigskip 

\textbf{Fig.1.} Schematic representation of the different patterns one finds
in the PDNLSE, Eq. (\ref{PDNLSE}). The line $\mu _{HB}$ has been ploted
after the numerical results of \cite{Bondila95}.

\textbf{Fig.2.} Amplitude of the bright cavity soliton, Gauss--Hermite mode (%
$GH_{1}$), and the mode $w_{2}$ (in dashed line).

\textbf{Fig.3.} Squeezing level (at the labeled frequencies) displayed by
the bright cavity soliton when a Gauss--Hermite mode ($GH_{1}$) or mode $%
w_{2}$ (see text) is used as a LOF. In (a) a $GH_{1}$ of width $\xi $ is
used. In (b) the LOF is displaced $x$ from the CS center. Parameters are $%
\Delta _{1}=1$, and $\mu =1.2$. 

\textbf{Fig.4.} Squeezing spectra close to the tangent bifurcation at $\mu
=1.0$. The detuning parameter is $\Delta _{1}=1.2$ and the pump values are
indicated in the figure. 

\textbf{Fig.5.} Imaginary parts of $w_{\mathrm{HB}+}^{\ast }$ (full line)
and $w_{\mathrm{HB}+}^{+}$ (dashed line). See text.

\textbf{Fig.6.} Squeezing spectrum at Hopf bifurcation. The dashed line
corresponds to using one of the two most damped modes (with $\func{Re}%
\lambda =0$) at this bifurcation as LOF. The full line corresponds to using
as LOF the sum of these two modes.

\textbf{Fig.7.} Squeezing spectra obtained when using a plane-wave LOF. The
detuning parameter is $\Delta _{1}=1.2$ and the pump value is indicated in
the figure. In the inset the maximum squeezing at $\omega =0$ is represented
as a function of pump for the same detuning value.

\textbf{Fig.8.} Squeezing spectra obtained when using a plane-wave LOF. The
pump parameter is $\mu =1.2$ and the detuning value is indicated in the
figure. In the inset the maximum squeezing at $\omega =0$ is represented as
a function of detuning for the same pump parameter value.

\textbf{Fig.9.} Spatial distribution of the squeezing when probed with a
plane-wave LOF and measured with finite size detectors of normalized size $%
\Sigma =\Delta x/\beta $ (the values are marked in the figure). The phase of
the LOF has been chosen in order to obtain the maximum noise redution at
each detector position. Parameters are $\mu =\Delta _{1}=1.2$.

\textbf{Fig.10.} Maximum squeezing $S_{out}$ at $\omega =0$, obtained with a
finite size detector centered at the BS center, as a function of the
detector size $\Sigma =\Delta x/\beta $. Parameters are $\mu =\Delta _{1}=1.2
$.


\begin{thebibliography}{99}
\bibitem{Drummond04} P. D. Drummond and Z. Ficek (eds.), \textit{Quantum
Squeezing} (Springer, 2004).

\bibitem{Loudon87} R. Loudon and P. L. Knight, J. Mod. Opt. \textbf{34}, 709
(1987).

\bibitem{Meystre91} P. Meystre and D. F. Walls (eds.), \textit{Nonclassical
Effects in Quantum Optics} (American Institute of Physics, New York, 1991).

\bibitem{Korolkova02} N. Korolkova, G. Leuchs, R. Loudon, T. C. Ralph, and
Ch. Silberhorn, Phys. Rev. A \textbf{65}, 052306 (2002).

\bibitem{Dodonov03} V. V. Dodonov and V. I. Man'ko (eds.), \textit{Theory of
Nonclassical States of Light}, (Taylor and Francis, London, 2003)

\bibitem{Walls94} D. F. Walls and G. J. Milburn, \textit{Quantum Optics},
(Springer Verlag, Berlin, 1994).

\bibitem{Reynaud87} S. Reynaud, C. Fabre, and E. Giacobino, J. Opt. Soc. Am.
B \textbf{4}, 1520 (1987).

\bibitem{Reid88} M. D. Reid and P. D. Drummond, Phys. Rev. Lett. \textbf{60}%
, 2371 (1988).

\bibitem{Drummond87} P. D. Drummond and S. J. Carter, J. Opt. Soc. Am. 
\textbf{4}, 1565 (1987).

\bibitem{Haus90} H. A. Haus and Y. Lai, J. Opt. Soc. Am. B \textbf{7}, 386
(1990).

\bibitem{Kozlov03} V. V. Kozlov, IEEE J. Sel. Topics Quant. Electron. 
\textbf{9}, 1468 (2003).

\bibitem{Treps00} N. Treps and C. Fabre, Phys. Rev. A \textbf{62}, 033816
(2000).

\bibitem{Lantz04} E. Lantz, T. Sylvestre, H. Maillotte, N. Treps, and C.
Fabre , J. Opt. B: Quantum Semiclass. Opt. \textbf{6}, S295 (2004).

\bibitem{Nagasako98} E. M. Nagasako, R. W. Boyd, and G. S. Agarwal, Opt.
Express \textbf{3}, 171 (1998).

\bibitem{Mecozzi98} A. Mecozzi and P. Kumar, Quantum Semiclass. Opt. \textbf{%
10}, 21 (1998).

\bibitem{Lugiato99} L. A. Lugiato, M. Brambilla, and A. Gatti, \textit{%
Optical Pattern Formation}, Adv. Atom. Mol. Opt. Phys. \textbf{40}, 229
(1999).

\bibitem{Lugiato02} L. A. Lugiato, A. Gatti, and E. Brambilla, J. Opt. B:
Quantum Semiclass. Opt. \textbf{4}, S176 (2002).

\bibitem{Kolobov99} M. I. Kolobov, Rev. Mod. Phys. \textbf{71}, 1539 (1999).

\bibitem{Lugiato95} L. A. Lugiato and G. Grynberg, Europhys. Lett. \textbf{29%
}, 675 (1995).

\bibitem{Lugiato93} L. A. Lugiato and A. Gatti, Phys. Rev. Lett. \textbf{70}%
, 3868 (1993).

\bibitem{Gatti95} A. Gatti and L. A. Lugiato, Phys. Rev. A \textbf{52}, 1675
(1995).

\bibitem{Drummond05} P. D. Drummond and K. Dechoum, Phys. Rev. Lett. \textbf{%
95}, 083601 (2005).

\bibitem{Zambrini00} R. Zambrini, M. Hoyuelos, A. Gatti, P. Colet, L. A.
Lugiato, and M. San Miguel, Phys. Rev. A \textbf{62}, 063801 (2002).

\bibitem{Gomila02} D. Gomila and P. Colet, Phys. Rev. E \textbf{66}, 046223
(2002).

\bibitem{EPL} I. P\'{e}rez-Arjona, E. Rold\'{a}n, and G. J. de Valc\'{a}%
rcel, Europhys. Lett. \textbf{74}, 247 (2006)

\bibitem{Gatti97} A. Gatti, H. Wiedemann, L. A. Lugiato, I. Marzoli, G. L.
Oppo, and S. M. Barnett, Phys. Rev. A \textbf{56}, 877 (1997).

\bibitem{Drummond80} P. D. Drummond and C. W. Gardiner,\ J. Phys. A \textbf{%
13}, 2353 (1980).

\bibitem{Gardiner00} C. W. Gardiner and P. Zoller, \textit{Quantum Noise}
(Springer, 2000).

\bibitem{Oppo94} G. L. Oppo, M. Brambilla, and L. A. Lugiato, Phys. Rev. A 
\textbf{49}, 2028 (1994).

\bibitem{Longhi97} S. Longhi,\ Phys. Scr. \textbf{56}, 611 (1997).

\bibitem{Trillo97} S. Trillo, M. Haelterman, and C. Sheppard, Opt. Lett. 
\textbf{22}, 970 (1997).

\bibitem{Longhi95} S. Longhi and A. Geraci, Appl. Phys. Lett. \textbf{67},
3060 (1995).

\bibitem{Bondila95} M. Bondila, I. V. Barashenkov, and M. M. Bogdan, Physica
D \textbf{87}, 314 (1995).

\bibitem{Barashenkov02} I. V. Barashenkov, N. V. Alexeeva, and E. V.
Zemlyanaya,\ Phys. Rev. Lett. \textbf{89}, 104101 (2002).

\bibitem{deValcarcel02} G. J. de Valc\'{a}rcel, I. P\'{e}rez-Arjona, and E.
Rold\'{a}n, Phys. Rev. Lett. \textbf{89}, 164101 (2002).

\bibitem{Collet85} M. J. Collet and D. F\ Walls, Phys. Rev. A \textbf{32},
2887 (1985).

\bibitem{Collet84} M. J. Collet and C. W. Gardiner, Phys. Rev. A \textbf{30}%
, 1386 (1984).

\bibitem{Alexeeva00} N. Alexeeva, I. V. Barashenkov, and G. P. Tsironis,
Phys. Rev. Lett. \textbf{84}, 3053 (2000).

\bibitem{Fauve90} S. Fauve and O. Thual,\ Phys. Rev. Lett. \textbf{64}, 282
(1990).

\bibitem{Miles84} J. W. Miles, J. Fluid Mech. \textbf{148}, 451 (1984).

\bibitem{Elphick89} C. Elphick and E. Meron, Phys. Rev. A \textbf{40}, 3226
(1989).

\bibitem{Barashenkov91} I. V. Barashenkov, M. M. Bogdan, and V. I. Korobov,
Europhys. Lett. \textbf{15}, 113 (1991).

\bibitem{Alexeeva99} N. V. Alexeeva, I. V.Barashenkov and D. E. Pelinovsky,
Nonlinearity \textbf{12}, 103 (1999).

\bibitem{Glauber63} R. J. Glauber, Phys. Rev. \textbf{130}, 2529 (1963); 
\textit{ibid.}\ \textbf{131}, 2766 (1963).

\bibitem{Gatti01} A. Gatti and S. Mancini, Phys. Rev. A \textbf{65}, 013816
(2001).
\end{thebibliography}
\end{document}